\documentclass[journal]{IEEEtran}
\usepackage{cite}

\ifCLASSINFOpdf
\else
\fi

\usepackage{amsmath}
\usepackage{amsfonts}
\usepackage[subtle]{savetrees}
\allowdisplaybreaks
\ifCLASSOPTIONcompsoc
 \usepackage[caption=false,font=normalsize,labelfont=sf,textfont=sf]{subfig}
\else
 \usepackage[caption=false,font=footnotesize]{subfig}
\fi

\usepackage{tikz}
\usetikzlibrary{calc,positioning}

\newtheorem{remark}{Remark}

\begin{document}

\title{Optimal Frequency Regulation using \\ Packetized Energy Management}

\author{Sarnaduti~Brahma,~\IEEEmembership{Member,~IEEE},
        Adil Khurram,~\IEEEmembership{Member,~IEEE}, \\
        ~Hamid~Ossareh,~\IEEEmembership{Senior Member,~IEEE}, and
        Mads~Almassalkhi,~\IEEEmembership{Senior Member,~IEEE}\thanks{This work was supported by the U.S. Department of Energy's Advanced Research Projects Agency-Energy (ARPA-E) award DE-AR0000694. M. Almassalkhi is co-founder of startup Packetized Energy, which seeks to bring to market a commercially viable version of Packetized Energy Management. The authors are with the Department of Electrical Engineering, The University of Vermont, Burlington,
        VT, 05405 USA. E-mail: \{\texttt{sbrahma,akhurram,hossareh,malmassa}\}@uvm.edu.}}

\maketitle

\begin{abstract}
Packetized energy management (PEM) is a demand dispatch scheme that can be used to provide ancillary services such as frequency regulation. In PEM, distributed energy resources (DERs) are granted uninterruptible access to the grid for a pre-specified time interval called the packet length. This results in a down ramp-limited response in PEM for DERs that can only consume power from the grid. In this work, a linearized virtual battery model of PEM is provided that is capable of predicting the down-ramp limited output of PEM and is used in a model predictive control (MPC) framework to improve the performance of PEM in tracking an automatic generation control (AGC) signal. By performing statistical analysis on the AGC regulation signal, PJM Reg-D, an ARMA model is derived as a predictor for the MPC-based precompensator. Finally, as an alternative to MPC, it is shown that by varying the packet length as a function of time, for example through packet randomization, frequency regulation can be improved under PEM.
\end{abstract}


\IEEEpeerreviewmaketitle
\IEEEpubid
\IEEEpubidadjcol

\section{Introduction}
The availability of connected and controllable distributed energy resources (DERs) has made it possible for grid operators to consider DERs in grid operations. Coordinating DERs in distributed control schemes engenders flexible demand that can deliver grid services while also ensuring the quality of service to the end-user~\cite{Morgan:1979, schweppe1980homeostatic}. The capability to provide services to the grid such as, but not limited to, peak-load reduction, energy arbitrage, and ancillary services with flexible demand is termed demand dispatch~\cite{Brooks:DemandDispatch, Callaway:2011wq}. This manuscript focuses on ancillary services, specifically, frequency regulation. 

Traditionally, frequency regulation or simply regulation is used to correct mismatches between load and supply by adjusting the power output of generators at fast time scales through automatic generation control (AGC). The regulation provided by generators is generally expensive. Aggregations of DERs, on the other hand, are a less expensive yet fast-acting alternative to generators that can provide regulation via demand dispatch~\cite{Kolter:PESGM2016}. Regulatory authorities such as Pennsylvania-New Jersey-Maryland interconnection (PJM)~\cite{PJM_AGC_url}, which is part of the Eastern Interconnection in the United States, generates an ``energy-neutral'' regulation AGC signal \textcolor{black}{(i.e., whose average over a sufficiently large period is almost zero)}, typically every two seconds, called Reg-D~\cite{pjmmanual}. This AGC signal is then transmitted to the DER coordinator, which in turn modifies the power output of DERs so that the aggregate power consumption tracks the AGC signal as accurately as possible. PJM has recently implemented an incentive-based `pay-for-performance' model that rewards resources that can provide regulation with high performance~\cite{PJM_pay4perf}. PJM measures performance using a combination of three metrics, Accuracy, Delay, and Precision, called Performance Score~\cite{pjmmanual}. \textcolor{black}{Here, the Accuracy score represents the maximum correlation between the input and output of the resource, taking into account a $10$ s delay. The Delay score represents the time delay at which correlation is the highest. Finally, the Precision score effectively represents the mean absolute tracking error between input and output.} The service provided by DERs via demand dispatch is also gauged using the same metrics.

Several demand-dispatch schemes have been proposed that aim to accurately track the AGC signal to meet regulatory requirements. These schemes generally differ in the amount of information needed to be transmitted between the coordinator and demand-side resources. Furthermore, the coordinator can be centralized, where a single entity generates ON/OFF commands and transmits them to the fleet~\cite{Callaway2009ECM, hao:tpwrst} or it can distributed, where each DER determines its control action depending upon local information and a common control signal from the coordinator~\cite{Mathieu:2013TPWRS,Luminita2014IFAC,abate:tcst, Meyn2015TAC, kundu:fitness, kundu:ai, hassan:privacyaware, emiliano:multi_period_OPF, Garcia:privacy_DER,kleissl:tsg, Lian:transactive, DuffautEspinosa:2020tpwrs, DuffautEspinosa:2020tcst,fathy:tcst,angstoch,tindeControl}. \textcolor{black}{For example, in \cite{angstoch}, the operating temperatures of refrigerators and their energy consumption, are modified dynamically,
within a safe range, in response to frequency fluctuations. In \cite{tindeControl}, a stochastic controller is developed to randomly switch TCLs to control the average power consumption.}

In distributed coordination, the aggregate dynamics of DERs are first modeled by discretizing a system of partial differential equations resulting in a state bin transition model~\cite{fathy:tcst,Mathieu:2013TPWRS,Luminita2014IFAC,abate:tcst,hassan:privacyaware, emiliano:multi_period_OPF}. Then, the control signal consisting of a vector of transition probabilities for DERs is obtained from control schemes such as the minimum variance controller~\cite{Callaway2009ECM}, model predictive control~\cite{Mathieu:2013TPWRS} or internal model control~\cite{fathy:tcst}. Each DER then measures its local state of charge (SoC) and switches its operating state according to the control input.  However, the vector of probabilities needs to be broadcast to all DERs which can be a communication burden. To avoid transmitting an entire vector, the concept of switching rate actuation is developed in~\cite{Luminita2014IFAC} that only requires a pair of turn-ON and turn-OFF probabilities. Similarly, the authors in~\cite{Meyn2015TAC} propose a scalar control signal that changes the DER's transition probabilities. 

In some cases, the information needed to generate the control input raises privacy concerns for the end-user. Therefore, privacy-aware dispatch has been proposed in~\cite{hassan:privacyaware} that uses differential privacy to protect the end-user's information. Privacy-preserving coordination has also been proposed in~\cite{Garcia:privacy_DER, kleissl:tsg} in which each DER determines its ON/OFF state locally using partial information from its neighbors. A fitness metric has been proposed in~\cite{kundu:fitness} in which DERs are dispatched depending upon their availability and quality of service they provide. In general, these schemes follow a top-down approach where a central coordinator or controller broadcasts the control signal to the entire fleet.

This paper focuses on a bottom-up device-driven architecture known as packetized energy management (PEM). In PEM, the DERs request the coordinator to consume power for a specified, fixed epoch, called packet length. The requests are either accepted or denied by the coordinator to regulate demand depending upon the provided power reference signal~\cite{Almassalkhi:2018IMA}. Packetized energy management has been developed and modeled in the author's earlier work~\cite{Almassalkhi:2018IMA,DuffautEspinosa:2018PSCC, DuffautEspinosa:2020tpwrs, DuffautEspinosa:2020tcst}. Specifically, population-based models and virtual battery (VB) models of PEM have been developed in~\cite{DuffautEspinosa:2020tpwrs} and~\cite{DuffautEspinosa:2020cdc}, respectively. Control architectures for diverse PEM fleets are presented in~\cite{DuffautEspinosa:2020tcst}. Whereas the population-based models require a large number of states, a VB model represents the dynamics of the fleet using a small number of key quantities, such as states for the average energy and aggregate power. That is, VB models are suitable for predicting the aggregate behavior of PEM for control applications such as model predictive control (MPC). 

In particular, one major challenge with PEM is that as packet requests are accepted by the coordinator, it locks devices ON for their packet length. This causes the aggregate response of DERs to become down ramp-limited. 
The down ramp-limited response is more prominent in DERs that can only consume power from the grid such as thermostatically controlled loads (TCLs) as the coordinator has no active mechanism through which it can reduce the load. That is, the aggregate power consumption of the TCL fleet can only decrease when a packet is completed and the TCL transitions to its OFF state. As a result, tracking error can increase when the reference signal decreases rapidly. The goal of this manuscript is to improve the tracking performance of PEM for fast frequency regulation, such as PJM's Reg-D.

In this paper, PEM is augmented with a novel MPC framework that enables DERs to provide fast grid services such as frequency regulation.  The focus is on TCLs such as electric water heaters (EWHs) that have a down-ramp limited response. An MPC-based precompensator is first proposed that adjusts the power reference signal based on the past information and future prediction of the fleet as well as the requested regulation service, AGC (e.g., PJM's Reg-D). \textcolor{black}{A challenge with this is that since PEM dynamics are essentially nonlinear, incorporating them as is into the MPC framework would lead to computational problems and possibly non-unique solutions. Hence,  PEM dynamics are modeled using a linearized virtual battery model, whereas the AGC signal is either assumed to be known (perfect forecast) or predicted using time-series forecasting.} PJM's performance scores as well as mean squared error (MSE) are used to gauge PEM's response to AGC. Simulation-based analysis has been conducted using the MPC-based precompensator together with an ARMA-model-based predictor derived through statistical analysis of the PJM regulation signal, Reg-D. Results show that the tracking performance of PEM improves significantly even in the case of the ARMA-based Reg-D forecast. 

In addition to an MPC precompensator, we also present a novel packet randomization scheme with which the tracking performance of PEM can be improved. The scheme allows the coordinator to accept packet requests and randomly select the packet length from a known distribution. Packet randomization engenders improved tracking during down-ramps by the availability of packets with shorter duration. Simulation results indicate that packet randomization has an added benefit of (slightly) reducing the number of times a TCL cycles between ON/OFF states, which can decrease wear-and-tear~\cite{prabir:cycl}. \textcolor{black}{Note that the packet randomization technique that is presented here is different from randomization approaches proposed in other papers. In this paper, the packet length (which is the duration for which DERs are allowed uninterrupted access to the grid) is randomized, whereas for example, in \cite{koch2011modeling}, the switching of DERs is randomized.}

The original contributions of this manuscript are as follows:
\begin{itemize}
    \item The aggregate power dynamics of PEM is described by a novel virtual battery model that captures the down-ramp limited nature of PEM. Specifically, the model adapts~\cite{DuffautEspinosa:2020cdc} to an MPC-ready model that is suitable for frequency regulation timescales. This is enabled by transforming the input from packet acceptance rate to power reference.
    \item An MPC-based precompensator is designed using the above model to improve tracking of PEM by  ``corner-cutting" the AGC, as we will show.
    \item A detailed statistical analysis of PJM Reg-D is provided to justify an ARMA model.
    \item Sensitivity studies are conducted with the MPC-based precompensator with respect to packet length and MPC prediction horizon, and also PJM performance scores. Specifically, the MPC framework enables PEM with a 5-minute packet length to outperform conventional PEM with a 3-minute packet length and improves PEM performance as the packet-length increases \textcolor{black}{(by as much as 10\% with a perfect forecast)}. This is significant as longer packet lengths lead to fewer device switches and a lower communication burden on the PEM scheme. \textcolor{black}{Moreover, there were no observed cases in these sensitivity studies where the performance deteriorates compared to the case where there is no MPC-based precompensator (i.e., the performance improvement is robust).}
    \item A new randomization mechanism is developed for PEM's packet length. Interestingly, using extensive numerical simulations, it is shown that employing randomized packet lengths improves both tracking performance  and reduces device cycling. 
\end{itemize}

The outline for the rest of the paper is as follows: Section \ref{sec:overview} provides a brief overview of Packetized Energy Management (PEM). Section \ref{sec:MPC} describes the PEM linearized VB model, develops the MPC-based precompensator and ARMA model for AGC, and reports simulation results with different packet lengths and MPC horizons. Section \ref{sec:tvpl} introduces time-varying packet length as an alternative method to improve tracking performance of PEM and investigates the cycling of devices. Finally, Section \ref{sec:Con} concludes the paper. 

\section{Overview of Packetized energy management for DERs}\label{sec:overview}
PEM for diverse DERs has been presented and modeled in the author's earlier work~\cite{Almassalkhi:2018IMA,DuffautEspinosa:2020tpwrs, DuffautEspinosa:2020tcst}. A brief description of PEM is presented next.

\begin{figure}[t]
    \centering
    \includegraphics[width=\columnwidth]{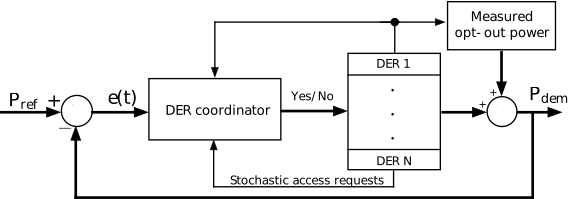}
    \caption{Closed-loop feedback system for PEM with $P_{\rm{ref}}$ provided by the grid or market operator and the aggregate net-load $P_{\rm{dem}}$ measured by the coordinator.}
    \label{fig:pem_schem}
\end{figure}

A DER with a local state-of-charge (SoC) $z[k]$, such as temperature for the case of electric water heaters, is designed to operate within a dead-band $[\underline{z}, \overline{z}]$. PEM's device-driven packet request mechanism is shown in the block diagram in Fig.~\ref{fig:pem_schem} and summarized below: 
\begin{itemize}
\item[$i)$] A DER measures or estimates its local SoC $z[k]$.
\item[$ii)$] If the SoC is within the dead-band, $z \in [\underline{z}, \overline{z}]$, the DER probabilistically requests the PEM coordinator to either consume power from the grid (charging) or inject power into the grid (discharging) for a pre-specified epoch and results in the notion of an energy packet. The epoch corresponding to the energy packet is called packet length and denoted as $\delta_\text{p}$. While the packet length was previously set to be a constant time interval (e.g., 5 minutes)~\cite{Almassalkhi:2018IMA}, this work introduces time-varying packet lengths. Irrespective of how the packet length is chosen, after a time equal to the packet length, has elapsed, the DER stops charging or discharging. 
\item[$iii)$] If the SoC is outside the deadband, $z \notin [\underline{z}, \overline{z}]$, the DER automatically and temporarily opts out of PEM to guarantee Quality of Service (QoS) and reverts to a conventional control mode until the SoC is returned within limits after which it returns to PEM operation.
\item[$iv)$] The PEM coordinator asynchronously accepts or denies the DER's packet request depending upon the reference tracking error. If the request is denied, go to $i)$; else, consume the energy packet and then go to $i)$.
\end{itemize}

Based on the previous description, a DER can be in either one of four modes; charging, discharging, OFF or opt-out. Furthermore, the probability that the DER with SoC $z[k] \in [\underline{z},\overline{z}]$ and desired set-point $z^\text{set}\in (\underline{z},  \overline{z})$ over time $k$ (for discretization time-step $\Delta t$) makes a request is given by a cumulative distribution function given by:
\begin{align}
P_{\text{req}}(z[k]) := 1 - e^{-\mu(z[k]) \Delta t},	\label{eq:preq}
\end{align}
where $\mu(z[k])>0$ is a rate parameter dependent on the local SoC. For \textit{charging} packet requests,
\begin{align}
\nonumber\lefteqn{\mu(z[k])} \\ 
& = 
\left\{\begin{array}{ll} 0, & \text{if $z[k] \ge \overline{z}$} \\
m_R \left(\frac{\overline{z} - z[k]}{z[k]-\underline{z}}\right) \cdot
\left(\frac{z^\text{set} - \underline{z}}{\overline{z} 
	-z^\text{set}}\right), & \text{if $z[k]\in (\underline{z}, 
	\overline{z})$}\\
\infty,  &  \text{if $z[k] \le \underline{z}$} \end{array}\right.,
\label{eq:PEM_MeanRequestRate}
\end{align}
where $m_R>0$ [Hz] is a design parameter that defines the mean time-to-request (MTTR) for $z=z^\text{set}$.
A similar expression follows for $\mu(z[k])$ and $P_{\text{req}}(z[k])$ in the case of \textit{discharging} packets.

\subsection{Aggregate power dynamics of PEM}
In this section, the virtual battery model of PEM developed in~\cite{DuffautEspinosa:2020cdc, khurram:phd} for aggregate power dynamics of PEM is adapted to obtain a model  that is suitable for the MPC formulation developed in the next section. The virtual battery model~\cite{DuffautEspinosa:2020cdc} for a fleet of DERs with state vector $x \in \mathbb{R}^{K}$ is:
\begin{align}
    x[k+1] &= f(x[k], u[k-T_d]), \quad y[k] = Cx[k]
\end{align}
where $f$ is a non-linear mapping, $f:\mathbb{R}^{K+1} \rightarrow \mathbb{R}^{K}$, $C\in \mathbb{R}^{1\times K}$ is a row vector, $y\in \mathbb{R}$ is the total power of the fleet, $u\in \mathbb{R}$ is the input ($P_\text{ref}$) to the virtual battery as shown in Fig.~\ref{fig:pem_schem}, and $T_d\ge 0 $ is the delay between the input  and output of PEM. This can account for communication delays between the system operator generating the AGC and PEM coordinator \textcolor{black}{and can be generally assumed to be constant, and the specific value can be determined from the regular operation of the system}. A delay is also assumed by PJM while computing the performance score \cite{pjmmanual} \textcolor{black}{that benefits resources unable to comply with the requirements}. Both the cases with ($T_d\ne 0$) and without delays ($T_d=0$) are considered. The state vector consists of $n_{\text{x}}$ states corresponding to aggregate quantities and $n_{\text{p}}$ states corresponding to the timer associated with packet requests, that is, $K = n_{\text{x}}+n_{\text{p}}$. The model of a DER fleet that can both charge and discharge consists of $n_{\text{x}}=4$ states, corresponding to aggregate quantities (SoC, charge, discharge and opt-out), and $n_{\text{p}}= \frac{\delta_{\text{p}} }{\Delta t}$ timer states. For DERs that can only charge, such as EWHs, $n_{\text{x}}=3$, and excludes the discharge state.
Thus, for a fleet of EWHs, the virtual battery model is given by:
\begin{align}
x_1[k+1] &= x_1[k](1 - \frac{\Delta t}{\tau}) + \frac{\Delta t x_{\text{amb}}}{\tau} \nonumber\\
&-\frac{1}{c \rho L}\left(Q - \frac{P_{\text{rate}}}{N}(x_2[k] + x_3[k]) \right), \label{eq:xsoc}\\ 
x_2[k+1] &= \frac{u[k-T_d]}{P_{\text{rate}}} - x_3[k] \label{eq:xon}\\
x_3[k+1] &= x_3[k](1-a_2) + a_1 P_{\text{req}}(x_1[k])(N -x_{\text{on}}[k]) \nonumber\\
&- a_1\left (\frac{u[k-T_d]}{P_{\text{rate}}} -x_{\text{on}}[k]\right), \label{eq:xopt}
\end{align}
where $x_1$ is the average SoC of the population, $x_2$ is the total number of EWHs whose requests have been accepted and are charging, and $x_3$ is the total number of EWHs that have opted out of PEM. Recall that in PEM, EWHs opt-out of PEM if their local SoC is outside of the allowed deadband (e.g., in the case of EWHs, this occurs when they get too cold or too hot). In the opt-out mode, EWHs charge without making any requests until their SoC is sufficiently recovered. In~\eqref{eq:xopt}, $P_{\text{req}}(x_1[k])$ models the total number of requests received by the coordinator. Furthermore, in~\eqref{eq:xsoc}, $P_{\text{rate}}$ is the rated power of the fleet, $Q$ is the average heat loss from the tank due to customer-driven water usage and is modeled as a Poisson random pulse process~\cite{DuffautEspinosa:2020tpwrs}, $\tau$ is the time constant related to heat-loss from the insulation of the water tank in seconds, $L$ is the tank size in liters, $c = 4.186$ $(\text{kJ})(\text{kg} ^\circ C)^{-1}$ is the specific heat constant and $\rho = 0.990$  $\text{kg}\text{L}^{-1}$ is the density of water when close to $50^\circ C$. The dynamics of the opt-out population is captured in~\eqref{eq:xopt} with parameters~$a_1$ and $a_2$~\cite{DuffautEspinosa:2020cdc}. For simplification, $x_{\text{on}}$ is introduced and is given by,
\begin{equation}\label{eq:xon1}
    x_{\text{on}}[k]:=x_{2}[k]+x_{3}[k]-z_{n_{\text{p}}}[k]
\end{equation} 
where $z_{n_{\text{p}}}$ accounts for the EWHs that have just finished their packets and are turning OFF. Then, $N - x_{\text{on}}[k]$ is the total number of EWHs in OFF mode and $P_\text{req}(x_1[k]) (N - x_{\text{on}}[k])$ models the proportion of EWHs that are making a request. Since EWHs can only charge for a time equal to $\delta_{\text{p}}$, $n_{\text{p}}$ timer states are added starting from $z_1,\dots, z_{n_{\text{p}}}$, to keep track of EWHs that have consumed a packet, resulting in the following simple dynamics,
\begin{align}
    z_1[k+1] &= \frac{u[k-T_d]}{P_{\text{rate}}}-x_{\text{on}}[k], \label{eq:z1}\\
    z_i[k+1]&= z_{i-1}[k], \quad \forall i = 2, \dots. n_{\text{p}},\label{eq:znp}
\end{align}
The full state vector is then $x := [x_1, x_2, x_3, z_1, \dots, z_{n_{\text{p}}}]^\top$ and $f$ in~\eqref{eq:xsoc} is given by the RHS of ~\eqref{eq:xsoc}-\eqref{eq:xopt},~\eqref{eq:z1}-\eqref{eq:znp}.

Note that~\eqref{eq:xon} contains the total number of EWHs that are charging at time $k$ and the state $z_1$ contains the newly accepted packets. Since the timer state, $z_1$, in~\eqref{eq:z1} cannot be negative, the following constraint applies to the control input $u[k]$ and captures the down-ramp limited nature of PEM, 
\begin{align}
    u[k-T_d] &\ge\!\! P_{\text{rate}}(x_{\text{on}}[k]).\label{ramp}
\end{align}
Similarly, $z_1$ cannot exceed the maximum number of packet requests received at time $k$, resulting in the following upper bound on $u[k]$, 
\begin{align}
    u[k-T_d] &\le\!  P_{\text{rate}} x_{\text{on}}[k] +   P_{\text{rate}} P_{\text{req}}(x_1[k])(N  - x_\text{on}[k])\label{ramp1}.
\end{align}
When tracking an AGC signal over short periods such as an hour, it is reasonable to assume that the AGC signal has been appropriately scaled so that~\eqref{ramp1} is generally not a binding constraint. This means that the coordinator receives a sufficient number of requests to enable upward ramps. Finally, the opt-out dynamics in~\eqref{eq:xopt} are nonlinear due to the probability of request~\eqref{eq:preq}. Furthermore, the AGC signal considered in this work is the Reg-D regulation signal provided by PJM, which is energy-neutral, as discussed in section~\ref{sec:arma}. Therefore, the state of charge $x_1$ is not expected to vary significantly over an hour and justifies the linearization of~\eqref{eq:xsoc}-\eqref{eq:xopt} for MPC as shown in the next section. 

\section{Model Predictive Control of Packetized Energy Management}\label{sec:MPC}
In this section, the PEM virtual battery model developed in the previous section is used to formulate an MPC to improve the tracking performance of PEM.
\subsection{Linearization of the VB model} \label{subsec:lin_vb}
In order to formulate the linear MPC problem, the virtual battery model~\eqref{eq:xsoc}-\eqref{eq:znp} is linearized as follows,
\begin{align}
    dx[k+1] &= (f_{0\;} - x_0) + Adx[k] + Bdu[k-T_d]\label{model1}\\
    y[k] &= y_0+dy[k]=C(x_{0}+dx[k])\\
    u_{0}+du[k-T_d] &\ge C_m (x_{0}+dx[k])\label{model2},
\end{align}
where $dx[k] = x[k] - x_0$, $du[k] = u[k]-u_0$, and $(x_0, u_0, y_0)$ is the nominal operating point, which is obtained by solving the nominal optimization problem discussed in~\cite{DuffautEspinosa:2020tpwrs}. Furthermore,  $A = \left.\frac{\partial f}{\partial x}\right|_{x_0,u_0}$ and $B = \left.\frac{\partial f}{\partial u}\right|_{x_0,u_0}$ are the Jacobians corresponding to the state and input respectively, and $C_m=[0\;1\;1\;0\;\cdots\;0\; -1]^{\top}$ as obtained from \eqref{eq:xon1} and \eqref{ramp}.

\subsection{MPC Formulation}
\begin{figure}[!t]
\centering
\begin{tikzpicture}[>=latex,scale=1, transform shape]
\node[draw] (mpc) at (0,0){MPC};
\node[draw] (pem) [right= of mpc]{PEM};
\node[draw] (pred) [left= of mpc]{Predictor};
\draw[->,thick] (pred)--node[above]{$R[k]$}(mpc);
\draw[->,thick] (mpc)--node[above]{$u[k]$}(pem);
\draw[->,thick] ($(pred)-(1.5,0)$)--node[above]{$r[k]$}(pred);
\draw[->,thick] (pem)--node[above]{$y[k]$}++(1.5,0)--++(0,-1)-|(mpc);
\draw[->,thick] ($(pem)+(1.5,0)$)--++(0.5,0);
\end{tikzpicture}
\caption{MPC-based Precompensator}
\label{fig:mpc}
\end{figure}
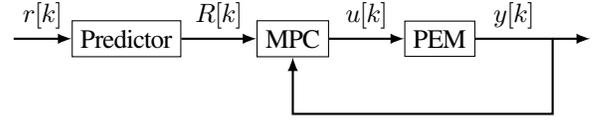

Consider a reference AGC signal $r[k]$ that is to be tracked by a fleet of DERs under PEM. The aim of this subsection is to design a precompensator that generates an optimal input $u[k]$ for PEM, such that its output $y[k]$ tracks $r[k]$. The proposed block diagram is shown in Fig.~\ref{fig:mpc}, and is based on a Model Predictive Control (MPC) approach. 

The linearized VB model developed in the previous subsection is utilized as the prediction model for the MPC. Assuming that the prediction horizon is $n$, the objective of MPC is to generate an optimal input trajectory $U[k]\in\mathbb{R}^n$ for the PEM model such that the tracking error between the model output $Y[k]\in\mathbb{R}^n$, and the reference trajectory $R[k]\in\mathbb{R}^n$ over the prediction horizon is minimized in some sense. Specifically in this paper, the focus is on two commonly used metrics, the 1-norm and the 2-norm of the tracking error, i.e., $\|Y[k]-R[k]\|_1$ and $\|Y[k]-R[k]\|_2$ respectively, as cost functions for the MPC. The 2-norm is a standard metric used in control theory, e.g., the linear quadratic regulator (LQR), while the 1-norm is used in the PJM precision score \cite{pjmmanual}.

Since the PJM precision score is defined to incentivize tracking of the past reference (which amounts to a delay), it is advantageous to utilize the $T_d$ previously occurred samples of the signal $r[k]$ as part of the reference trajectory $R[k]$ for MPC, and predict the remaining part of $R[k]$, containing $n-T_d$ samples, using a predictor, for example, an ARMA model. Thus,  $R[k]=[r[k-T_d+1],\;r[k-T_d+2],\;..., \; r[k],\; r^*[k+1],\; ...\;r^*[k-T_d+n]]^{\top}$, where $r^*[k]$ are predicted values of the reference in the future.
 Using \eqref{model1}-\eqref{model2}, the MPC problem can then be formulated as:
\begin{equation}\label{mpcprob}
    \begin{array}{cc}
\textrm{minimize} & \left\Vert Y_{0}+dY-R\right\Vert _{p}^p\\
\textrm{over }dx,\;du\\
\textrm{subject to:} & dY-M_{y}dU=G_{y}\\
 & M_{u}dU\preceq G_{u1}-G_{u2}
\end{array}
\end{equation}
where:
\begin{align*}
Y_0 &= [1\quad\cdots\quad 1]^\top y_0\\
    dY	&=\left[\begin{array}{c}
Cd{x}\left[k+T_d+1\right]\\
\vdots\\
Cd{x}\left[k+T_d+n\right]
\end{array}\right],\quad dU=\left[\begin{array}{c}
d{u}\left[k\right]\\
\vdots\\
d{u}\left[k+n-1\right]
\end{array}\right]\\
M_y&=\left[\begin{array}{cccc}
CB & 0 & \cdots & 0\\
CAB & CB & \ddots & \vdots\\
\vdots & \ddots & \ddots & 0\\
CA^{n-1}B & \cdots & CAB & CB
\end{array}\right]\\
G_y &=\left[\begin{array}{c}
C\left(f_{0\;}-x_{0}\right)\\
C\left(f_{0\;}-x_{0}+A\left(f_{0\;}-x_{0}\right)\right)\\
\vdots\\
C\left(f_{0\;}-x_{0}+\cdots+A^{n-1}\left(f_{0\;}-x_{0}\right)\right)
\end{array}\right]\\
M_u&=\left[\begin{array}{cccc}
-1 & 0 & \cdots & 0\\
C_{m}B\; & -1 & \ddots & 0\\
\vdots\; & \ddots\; & \ddots & \vdots\\
{C_{m}\;A}^{n-2}B & \cdots & C_{m}B & -1
\end{array}\right]\\
G_{u1}&=\left\lbrack \begin{array}{c}
1\\
\vdots\\
1
\end{array}\right\rbrack \left(u_{0}-C_{m}x_{0}\right)
\end{align*}
\begin{equation*}
    G_{u2}=\left\lbrack \begin{array}{c}
0\\
C_{m}\left(f_{0\;}-x_{0}\right)\\
C_{m}\left(f_{0\;}-x_{0}+A\left(f_{0\;}-x_{0}\right)\right)\\
\vdots\\
C_{m}\left(f_{0\;}-x_{0}+\cdots+A^{n-2}\left(f_{0\;}-x_{0}\right)\right)
\end{array}\right\rbrack
\end{equation*}
and $p=1$ (linear objective) or 2 (quadratic objective). Note that the Jacobians in Section \ref{subsec:lin_vb} are re-linearized \textcolor{black}{(e.g., the Hessians are recomputed)} at every time step of the simulation using observed system data, and the MPC states re-initialized at the new nominal operating point, $(x_0, u_0, y_0)$. Solving the MPC problem~\eqref{mpcprob} results in the optimal control input $U[k]$, only the first element of which is applied at every time step as the input to PEM. Furthermore, in simulations presented in this paper, it is assumed that the state vector $x[k]$ is measured. \textcolor{black}{Specifically, apart from the reference signal, the MPC needs to have access to estimates of the mean temperature (SoC) of the devices, the number of devices that are charging and have opted out, as well as the timer states. In an earlier work \cite{espinosa2020virtual}, we showed that the nonlinear  VB model is strong locally observable, which enables construction of an Extended Kalman Filter (EKF) to accurately provide these estimates.}
 In the subsequent time steps, the total number of charging population and the deterministic timer dynamics follow directly from the control input $U[k]$ and the total number of opt-out EWHs are provided by PEM. However, the SoC may not be measured but can be obtained from the VB model. This is because SoC is not expected to vary significantly over an hour provided an accurate initial estimate. This completes the description of MPC for PEM. The next subsection presents an example predictor of AGC for MPC, obtained using the ARMA modeling of the AGC signal, PJM Reg-D.

\subsection{PJM Reg-D statistics and ARMA modeling}\label{sec:arma}
As shown in the previous subsection, we need a prediction model for the AGC signal to generate a prediction of AGC over a horizon that can be used by the MPC-based precompensator to generate the optimal modified AGC. In this subsection, we provide an example forecasting model for a commonly used AGC signal that can be used for the MPC-based precompensator. The AGC regulation signal used in this paper is a scaled and biased version of Reg-D, which is one of the regulation signals provided by PJM \cite{pjmmanual} (since Reg-D is normalized between $-1$ and $1$, $1$ representing the maximum and $-1$ the minimum power bid into the market, it needs to be scaled and biased by the appropriate power to generate the actual AGC signal). Compared to the Reg-A, which is another, slower, PJM regulation signal that is meant to recover larger, longer fluctuations in system conditions, the Reg-D is a fast, dynamic signal whose hourly average tends towards zero (i.e., it is \textit{energy-neutral}) but requires resources to respond rapidly~\cite{pjmmanual}.

\begin{figure}[t]
\vspace{-12pt}
\centering
\subfloat[Autocorrelation]{\includegraphics[width=0.48\columnwidth]{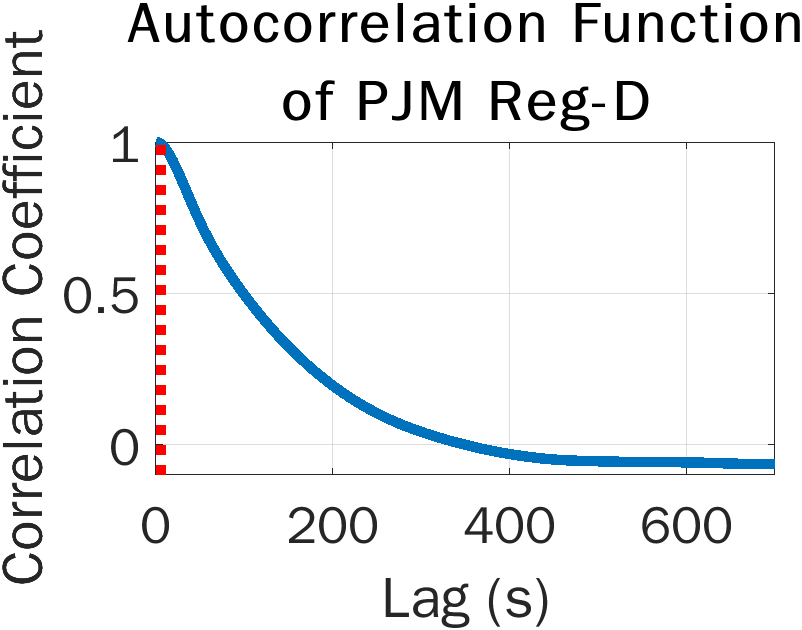}\label{fig:corrcoef}}
\hfil
\subfloat[Partial Correlation]{\includegraphics[width=0.5\columnwidth]{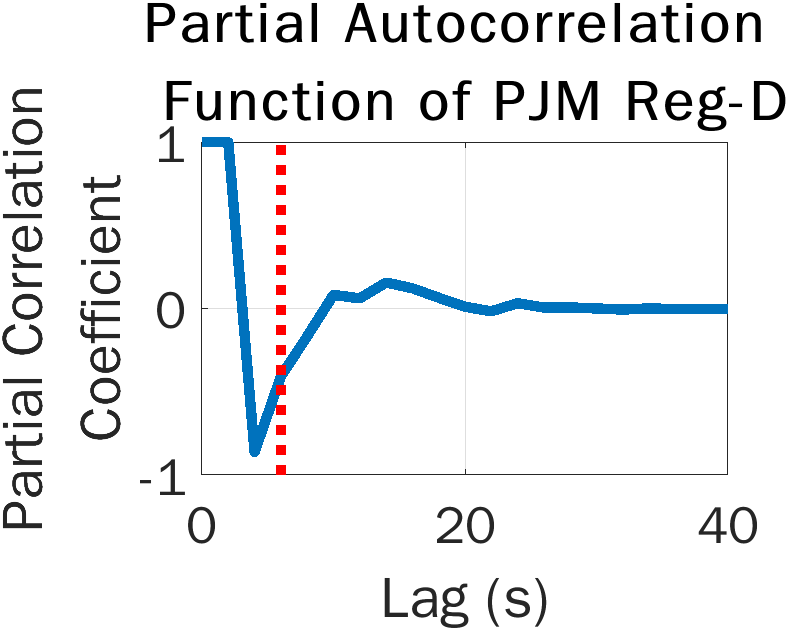}\label{fig:parcorr}}
\caption{Autocorrelation and Autocorrelation functions of PJM Reg-D. The lag considered for the ARMA model is shown by the red vertical dotted line.}
\label{agccorr}
\end{figure}

\begin{figure}[t]
\centering
\includegraphics[width=0.6\columnwidth]{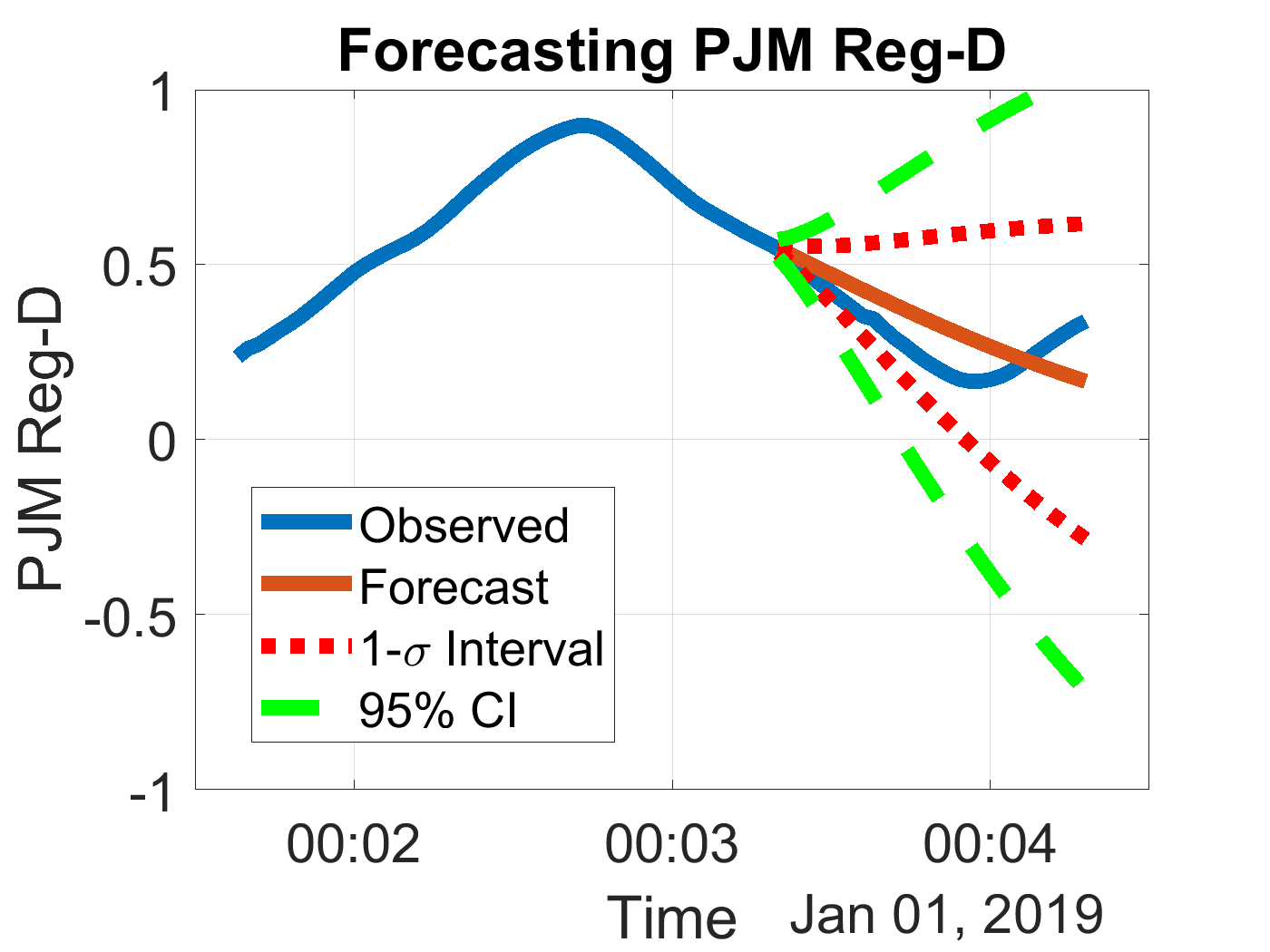}
\caption{Forecasting Reg-D. Times are in hh:mm format.}
\label{fig:forecast}
\vspace{-10pt}
\end{figure}

\begin{figure}[t]
\vspace{-10pt}
\centering
\subfloat[Minutely Variance]{\includegraphics[width=0.48\columnwidth]{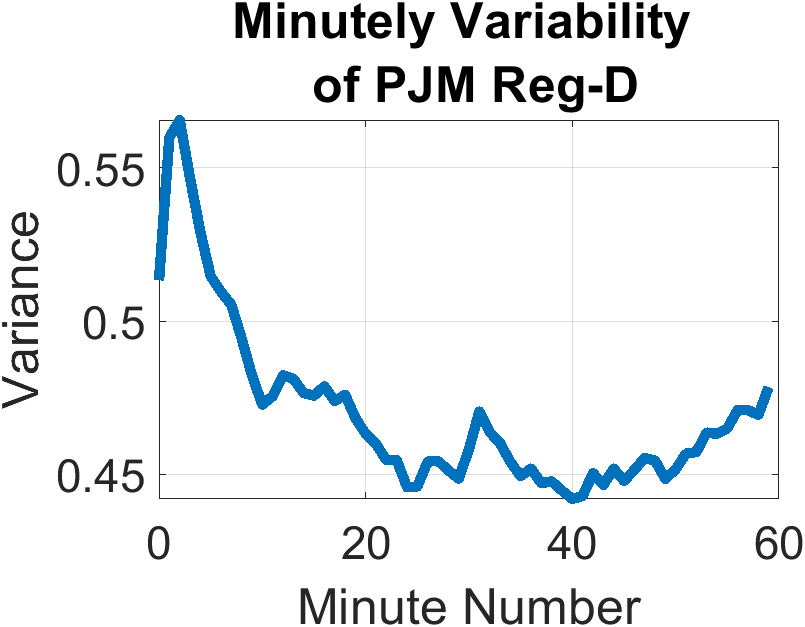}\label{fig:varmin}}
\hfil
\subfloat[Hourly Variance]{\includegraphics[width=0.48\columnwidth]{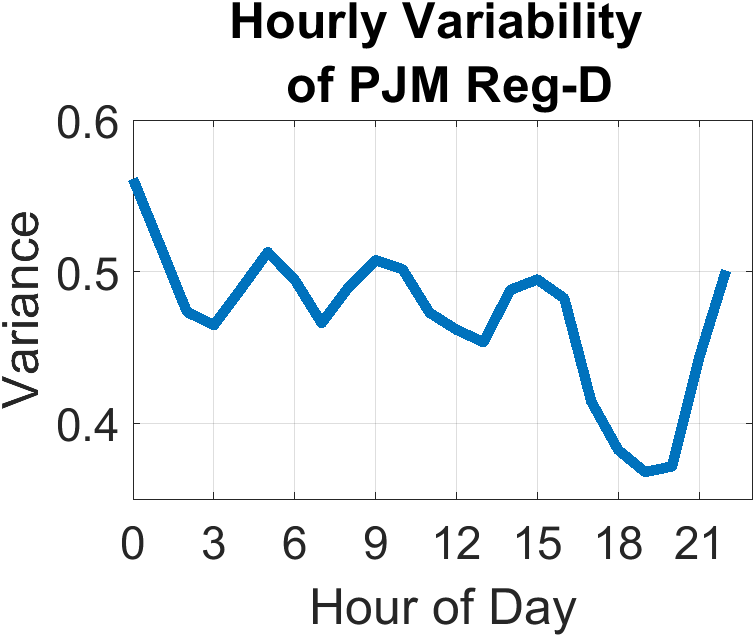}\label{fig:varhour}}
\hfil
\subfloat[Daily Variance]{\includegraphics[width=0.47\columnwidth]{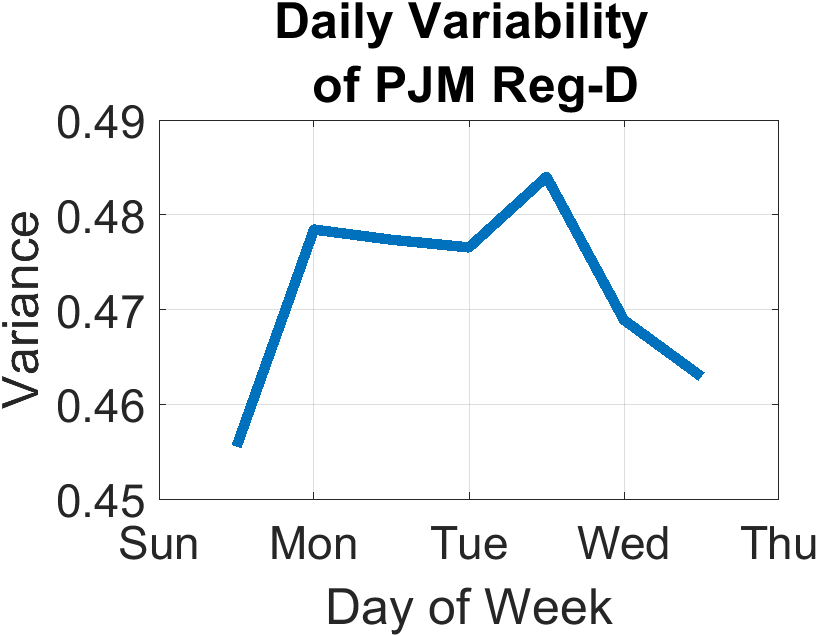}\label{fig:varday}}
\subfloat[Monthly Variance]{\includegraphics[width=0.52\columnwidth]{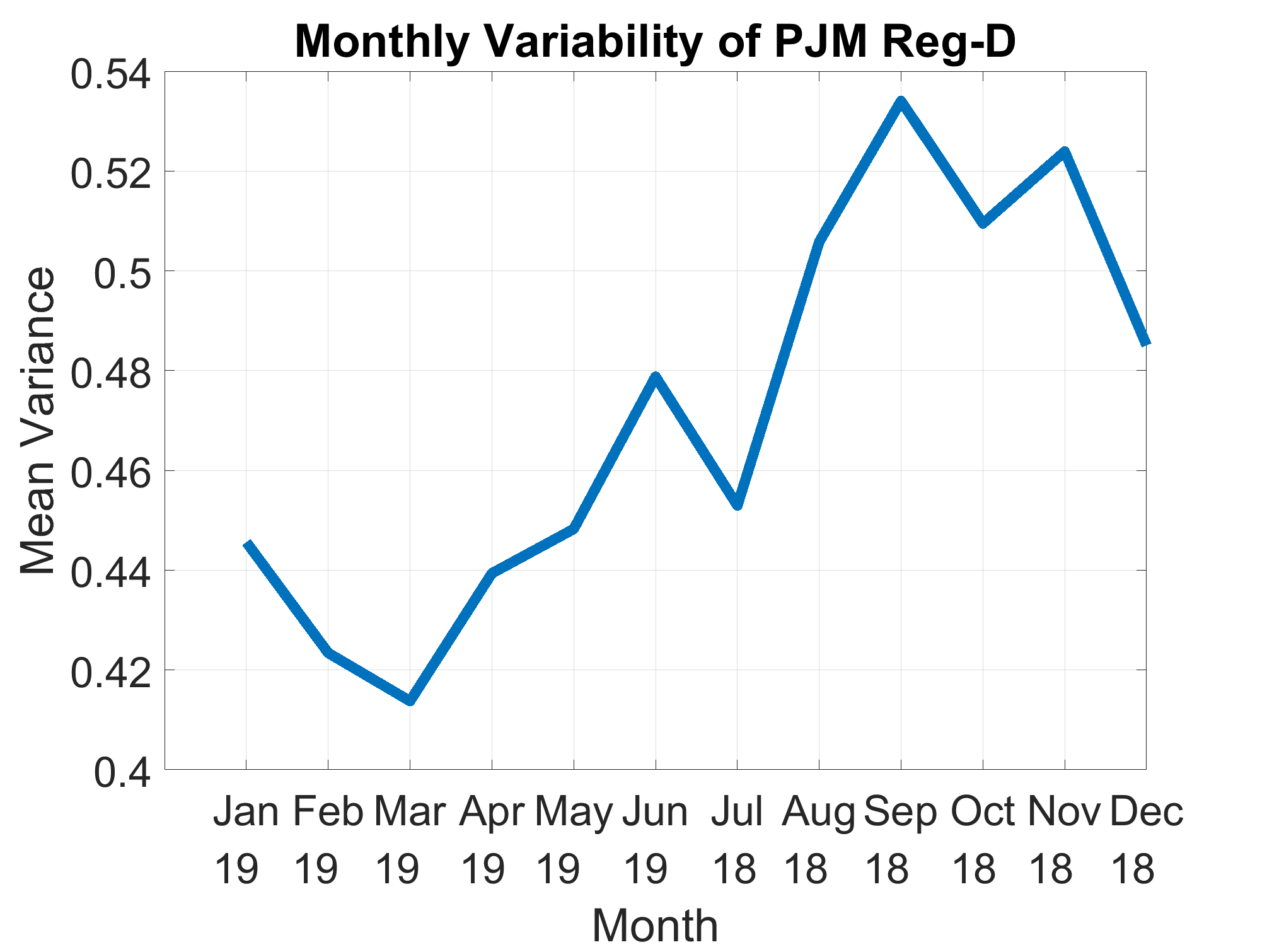}\label{fig:varmonth}}
\caption{Variability of PJM Reg-D}
\label{regdvar}
\end{figure}
The prediction model for the AGC is assumed to be an autoregressive moving average (ARMA) model \cite{box2015time}:
\begin{align}
     r[k]={}&\mu+\phi_{1}r[k-1]+\ldots+\phi_{p}r\left[k-g\right]\nonumber\\
     &+a\left[k\right]-\theta_{1}a\left[k-1\right]-\ldots-\theta_{q}a\left[k-h\right]
\end{align} where $r[k]$ is the AGC time-series, $\phi_i$ are the autoregressive coefficients,  $\theta_i$ are the moving-average coefficients, $a[k]$ is a standard Gaussian white noise process, and $\mu$ is the mean level of the ARMA process. To find the order $(g,h)$ of the ARMA model, the autocorrelation function (ACF) and the \textcolor{black}{partial autocorrelation function (PACF)} are plotted for PJM Reg-D from January 2019. They are shown in Fig.~\ref{agccorr}. \textcolor{black}{The ACF of a stationary real stochastic process $r[k]$ at a lag $l$ is defined as: $\rho_l=E((r[k-l]-\mu)(r[k]-\mu))$. The PACF $\Phi_{ll}$ of $r[k]$ at a lag $l$ is defined as the ``partial correlation" between $r[k]$ and $r[k-l]$, which is the correlation between $r[k]$ and $r[k-l]$ that is not accounted for by the intermediate values $r[k-l+1]$, $r[k-l+2]$, ..., $r[k-1]$, and is computed by solving:
\[\left[\begin{array}{cccc}
1 & \rho_{1} & \cdots & \rho_{l-1}\\
\rho_{1} & 1 & \ddots & \vdots\\
\vdots & \ddots & \ddots & \rho_{1}\\
\rho_{l-1} & \cdots & \rho_{1} & 1
\end{array}\right]\left[\begin{array}{c}
\phi_{l1}\\
\phi_{l2}\\
\vdots\\
\phi_{ll}
\end{array}\right]=\left[\begin{array}{c}
\rho_{1}\\
\rho_{2}\\
\vdots\\
\rho_{l}
\end{array}\right]\]
where $\rho_i$ is the autoregressive coefficient at lag $i$. The PACF is often encountered in the literature of system identification. Details can be found in \cite{pfeifer1980identification,akaike1979bayesian}.} It can be seen that the ACF tails off (and does not cut off), while the PACF cuts off after about 3 lags (or 6 s, for a sample time of 2 s). This suggests an autoregressive model of order 3 (i.e., AR(3)) or more for the AGC signal. In this study, the lag of 3 time steps is considered on grounds of parsimony. The autoregressive coefficients can then be selected using, e.g., Yule-Walker equations~\cite{box2015time}, to fit historical Reg-D data. Fig. \ref{fig:forecast} shows a snapshot of PJM Reg-D being forecasted using an AR(3) model. It can be seen that the actual signal is within 1 standard deviation of the mean prediction. From studies on predicting PJM Reg-D signals using AR(3) model, it is found that the error in predicting the value of Reg-D 1-min ahead is less than 15\%.

To ascertain the validity of this model for different segments of AGC, the variability of the Reg-D signal was investigated on a minutely, hourly, daily and monthly basis. Specifically, the mean variance of Reg-D across different minutes of the hour, hours of the day, days of the week, and months of the year from July 2018 to June 2019 were evaluated. The results are shown in Fig. \ref{regdvar}. Fig. \ref{fig:varmin} shows that there is a peak in variability at the beginning of every hour. This can correspond to changes in the electricity market every hour, resulting in step changes in generator set-points and consequently in  power flows. Fig. \ref{fig:varhour} shows that Reg-D changes rapidly from being the least variable at 7 pm (after the sun sets) to the most variable at midnight. The daily variability is fairly constant, as can be seen from Fig. \ref{fig:varday}. The monthly variability of Reg-D (Fig. \ref{fig:varmonth}) changes from the lowest before summer (around March) to the highest after summer (around September). From all the subfigures of Fig. \ref{regdvar}, it can be seen that the change in  variability of Reg-D is less than 10\% of its range, which is equal to 2 (-1 to 1). This provides confidence that the ARMA model is effective in predicting the AGC over multiple time segments.

\subsection{Simulations with No Time Delay}\label{mpcsim}
Simulations were conducted on PEM, equipped with the MPC-based precompensator, first with $T_d=0$, to test its performance for different packet lengths. To take into account a variety of AGC signals \textcolor{black}{that are representative of different times of the day and seasons of the year}, twelve 1-h PJM Reg-D signals from twelve different months and times of the day, starting from July 2018 and ending in June 2019, were taken. The PJM Reg-D signals, ranging from -1 to 1, were scaled by 1 MW around the nominal power consumption of the fleet of 6000 DERs, which is 3.7 MW.

Hence, AGC signals with a maximum value of $r_\textrm{max}=4.7$ MW and a minimum value $r_{\textrm{min}}=2.7$ MW were considered for the tests. The sample time of the simulations was taken to be 2 s, which is the same as that of the AGC data. This is acceptable since the discretization errors are not significant.

\subsubsection{Linear Objective Function}\label{linobjres}
\begin{figure}[t]
\centering
\subfloat[With Linear Cost]{\includegraphics[width=0.48\columnwidth]{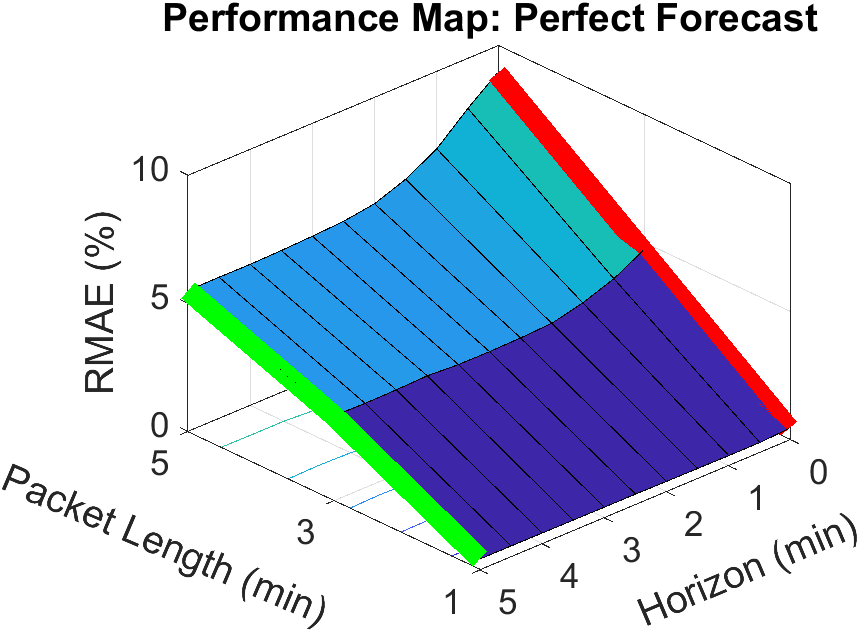}\label{fig:pmaplp}}
\hfil
\subfloat[With Quadratic Cost]{\includegraphics[width=0.48\columnwidth]{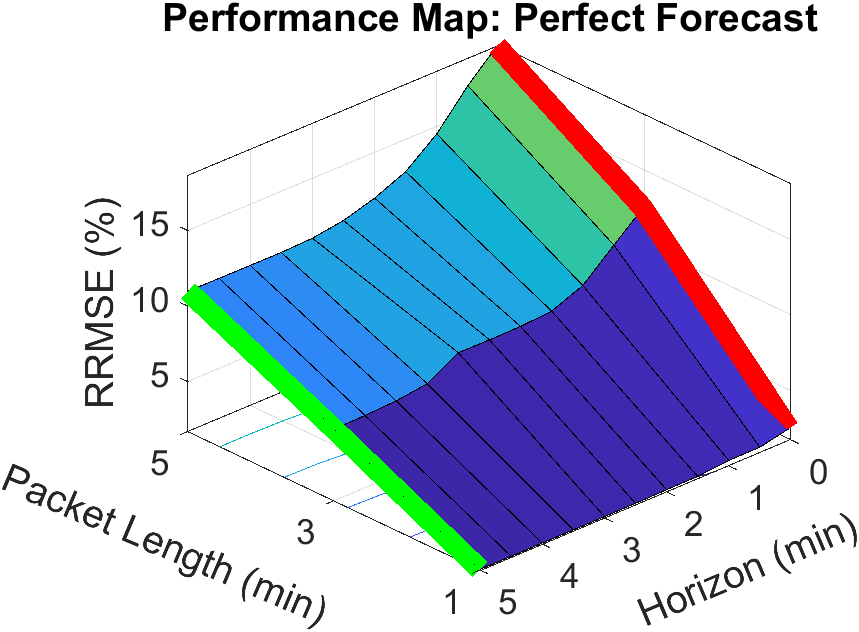}\label{fig:pmapqp}}
\caption{Performance Map of MPC with perfect forcast}
\label{pmappf}
\end{figure}

The results for $p=1$ in \eqref{mpcprob} are shown in Figs. \ref{fig:pmaplp}, assuming a perfect forecast for MPC. The y-axis plots the relative mean absolute error, defined as:
\begin{equation}
    \textrm{RMAE}=\frac{1}{12}\sum_{i=1}^{12}\left(\frac{\sum_{k=1}^{m}\left|y_{i}\left[k\right]-r_{i}\left[k\right]\right|}{m\left(r_{\textrm{max}}-r_{\textrm{min}}\right)}\right),
\end{equation}
where $r_i$ and $y_i$ refer to the $i$th AGC reference signal (out of the 12 signals considered around the year) and the corresponding PEM output signal respectively. $m$ is the number of samples present in the AGC signal in one hour. It can be seen from Fig. \ref{fig:pmaplp} that with a perfect forecast, there is a performance improvement of up to 3.5\% for a 5 min packet length (consider the red and green lines). With AR(3) forecast, the corresponding performance improvement reduces to 1.1\%.

\begin{figure}[t]
\centering
\subfloat[Intersection at Down Ramps caused by MPC]{\includegraphics[width=0.48\columnwidth]{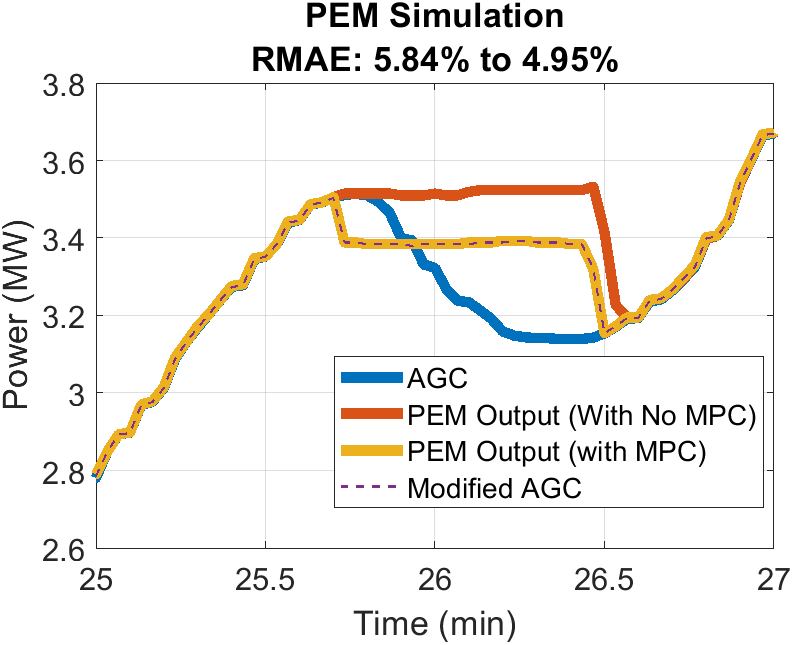}\label{fig:mpccut}}
\hfil
\subfloat[Effect of AGC range on MPC performance]{\includegraphics[width=0.48\columnwidth]{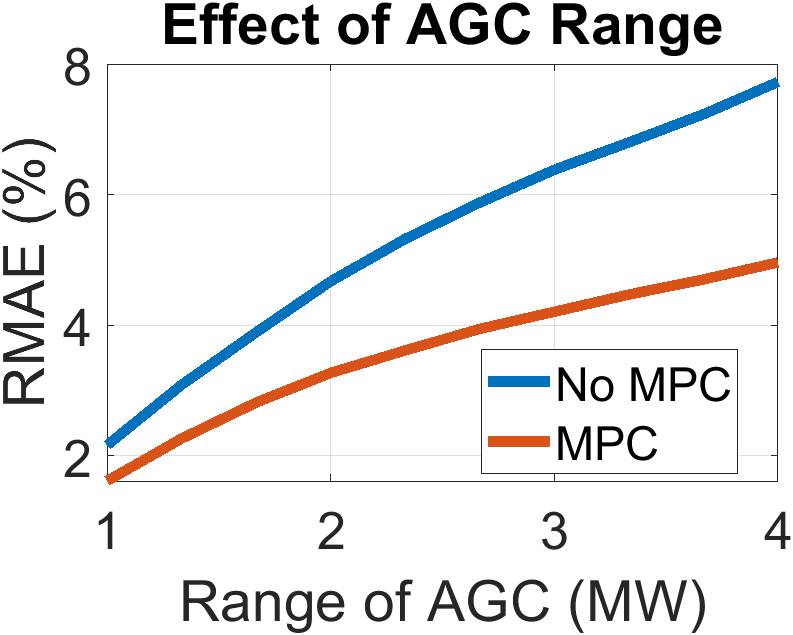}\label{fig:agcrange}}
\caption{Performance MPC-based precompensator}
\end{figure}

\subsubsection{Quadratic Objective Function}
Similarly, the results for $p=2$ in \eqref{mpcprob} are shown in Fig. \ref{fig:pmapqp} for a perfect forecast. The y-axis reports the relative root mean squared error:
\begin{equation}
    \textrm{RRMSE}=\frac{1}{12}\sum_{i=1}^{12}\left(\frac{\sqrt{\frac{1}{m}\sum_{k=1}^{m}\left(y_{i}\left[k\right]-r_{i}\left[k\right]\right)^{2}}}{r_{\textrm{max}}-r_{\textrm{min}}}\right)
\end{equation}
where $r_i$, $y_i$, $m$, $r_{\textrm{max}}$, and $r_{\textrm{min}}$ are the same as in Section \ref{linobjres}. It can be seen from Fig. \ref{fig:pmapqp} that with a perfect forecast, there is a performance improvement of up to 9.8\% for a 5 min packet length, which is much larger than that obtained using a linear cost function. This makes sense since the quadratic cost function penalizes large errors heavily. With AR(3) forecast, the corresponding performance improvement was lesser (as expected): around 2\%, but still higher than the corresponding case with a linear objective function. 

The above analyses indicate that an accurate forecast can significantly improve tracking performance. Specifically, with a perfect forecast, we can obtain up to 4\% improvement in RMAE and 10\% improvement in RRMSE with a 5 min packet length. While an ARMA forecasting model is described here to illustrate the method, future work includes the development of improved forecasting tools (e.g., using neural networks, VARMA models).

\subsubsection{Intersection at Down-ramps}
The improvements due to MPC are mainly because of the intersections between the AGC and the MPC output occurring at down-ramps, which is illustrated in Fig. \ref{fig:mpccut}. Since the output of PEM is down-ramp limited, if it can be predicted when a down-ramp event will occur, the MPC-based precompensator can instruct PEM to track the down-ramp ahead of time, resulting in the PEM output intersecting the AGC as it ramps down. Hence, essentially, the predictor anticipates when the AGC down-ramps and MPC uses that information to precompensate the AGC for that down-ramp.

\subsubsection{Effect of AGC range}
To quantify the effect of AGC range on MPC tracking performance, simulations were conducted on the same twelve 1-h AGC samples but scaled accordingly such that they ranged from 1 to 4 MW. The MPC horizon was fixed to be 3 min and the packet length was also considered to be 3 min. The size of the fleet was kept to be the same as before, namely 6000 DERs, and was held constant. The effect of the range of AGC is illustrated in Fig.~\ref{fig:agcrange} for a perfect forecast. It can be seen that the performance improvements with MPC are larger when the range of AGC increases. This is expected because when the range is small, PEM is less down-ramp limited and hence itself tracks well so that the performance improvement with MPC is smaller in such a case.

\subsubsection{Run Time}
The MPC optimization takes less than 0.5~s to be solved. Specifically, using Gurobi solver \cite{gurobi} on a laptop with a 2.2 GHz processor, it took around 400 ms at the worst-case time-step for 5 min packet length. With packet lengths equal to 3 min, the worst-case time was far smaller, less than 50 ms.

\subsection{Simulations with Time Delay: PJM Performance Scores}\label{sec:pjm}
To quantify the performance of a regulating resource in tracking a dispatch signal, PJM utilizes a ``Performance Scoring" mechanism \cite{pjmmanual}, on a scale from 0 to 1. This scoring currently incorporates a 10-s delay between the input and output of the resource, arising, e.g., due to communication delays. The PJM composite score is the average of three scores: the accuracy, precision, and delay scores. The accuracy score represents the maximum correlation between the input and output of the resource, taking into account the $10$ s delay. The delay score represents the time delay at which correlation is the highest. Finally, the precision score effectively represents the mean absolute tracking error. The definitions of these scores are reviewed in detail in the Appendix since the authors were unable to find an archive of these formulae in the literature.

Since the PJM scores consider a $10$ s delay between the input and output, in the formulation \eqref{mpcprob}, a value of $T_d=5$  (for a sample time of 2 s) is used, as opposed to $T_d=0$ in the previous subsection. The value of $p$ is taken to be $1$ in this case since the PJM precision score uses a term that involves (but is not exactly) the mean absolute error between the input and the output, and not the mean squared error. Furthermore, it should be mentioned here that the exact PJM precision score \cite{pjmxls} leads to a nonconvex cost function. However, investigation of the exact form of the PJM precision score is out of the scope of this paper and is deferred to future publications. The same simulation scenarios as in the previous subsection are considered next. Moreover, a delay-based precompensator is also attempted that essentially delays the AGC by $10$ s before sending it to PEM. 
\begin{figure}[t]
\centering
\subfloat[Packet Length: 1 min]{\includegraphics[width=0.48\columnwidth]{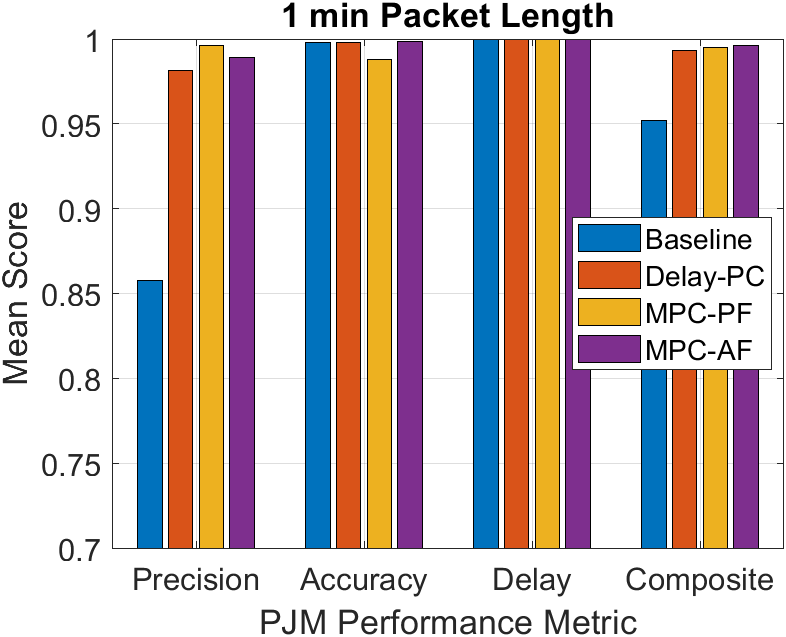}\label{fig:pl1min}}
\hfil
\subfloat[Packet Length: 3 min]{\includegraphics[width=0.48\columnwidth]{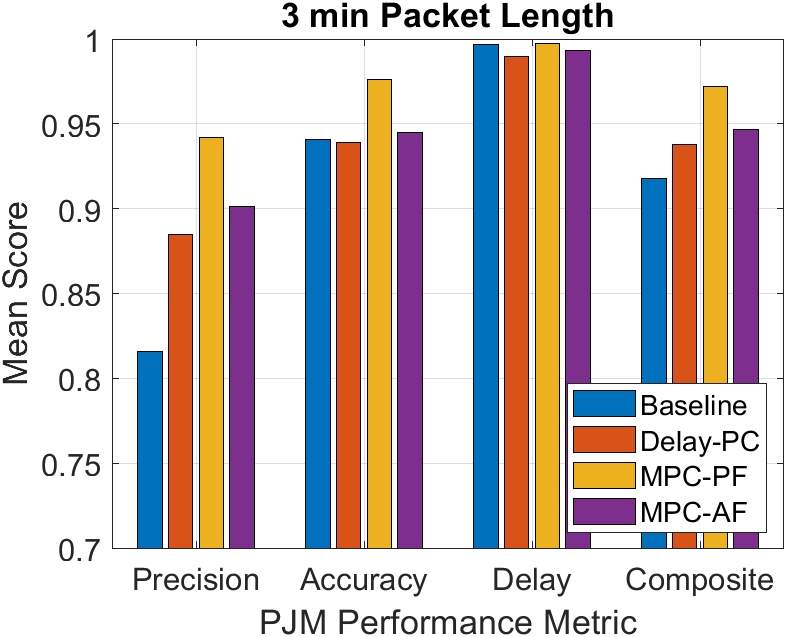}\label{fig:pl3min}}
\subfloat[Packet Length: 5 min]{\includegraphics[width=0.48\columnwidth]{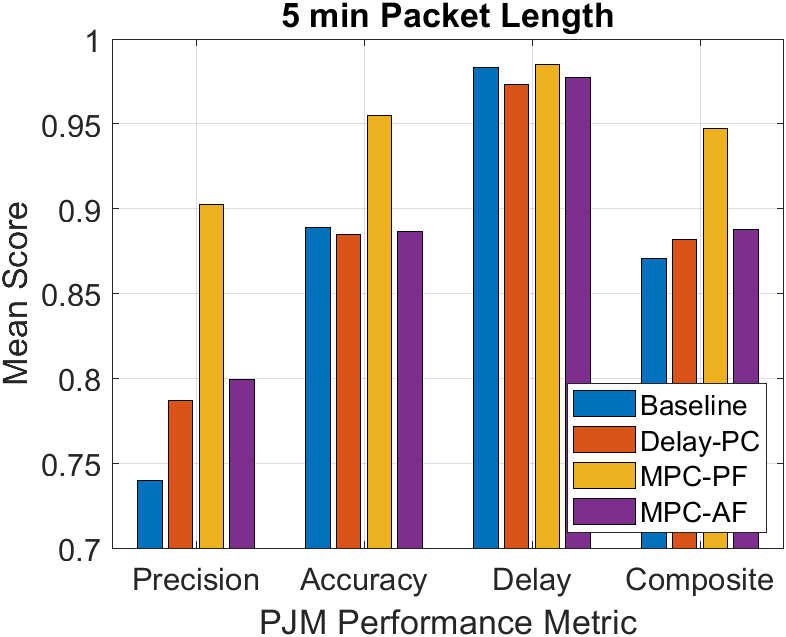}\label{fig:pl5min}}
\caption{Comparision of PJM Performance Scores for different methods. The legends describe different input signals provided to PEM, generated with/without a precompensator. ``Precision", ``Accuracy", ``Delay" and ``Composite" scores are different performance metrics.}
\label{pjmscores}
\end{figure}
Fig. \ref{pjmscores} shows the results from all the methods. In this figure, ``Baseline" stands for raw AGC (i.e., the shifted and scaled PJM Reg-D), ``Delay-PC" stands for a $10$ s delayed AGC generated using the delay-based precompensator, ``MPC-PF" stands for AGC passed through the MPC-based precompensator using a perfect forecast, and ``MPC-AF" stands for that using AR(3) forecast, as the respective inputs to PEM.

It can be seen in Fig.~\ref{pjmscores} that if the forecasts are perfect, the PJM precision score increases by up to $10\%$ for $5$ min packet length over that obtained using a delay-based precompensator and up to $15$\% over that obtained with no precompensator. For the same packet length but with an AR(3) forecast, the performance improvement is about $1$\% over that obtained using a delay-based precompensator, and about $5$\% over that obtained using no precompensator. Similar trends can be observed in the PJM composite score as well. Moreover, with AR(3) forecast, we also obtain robust performance improvement \textcolor{black}{(specifically, there were no observed cases in which the performance of the MPC-based precompensator deteriorated compared to the case when there is no precompensator).} These observations indicate that having a better forecast will improve the performance of the precompensator significantly (the investigation of which, as mentioned earlier, is a topic of future research). Furthermore, it is observed that MPC can make PEM with 5 min packet length achieve similar or better performance as that of a 3 min packet length, and 3 min as that of PEM with 1 min packet length. 

\section{Time-varying packet length in PEM}\label{sec:tvpl}
The previous section presented an MPC formulation to improve the tracking performance of PEM when tracking an AGC signal by the intersection at down-ramps. This section shows that tracking performance in PEM can also be improved by varying the packet length. First, the effect of different packet lengths on tracking error is quantified. Then, a time-varying packet length scheme is proposed for the PEM system that improves tracking performance.
\subsection{Effect of packet length}
In PEM, shorter packet lengths allow for better tracking performance than longer packets. The reason is that PEM has a down ramp-limited response, as discussed in the previous section. Shorter packets expire faster and allow the coordinator to make adjustments to the accepted requests so that tracking error is minimized. 

\begin{figure}[t]
\vspace{-4pt}
    \centering
    \includegraphics[width=0.9\columnwidth]{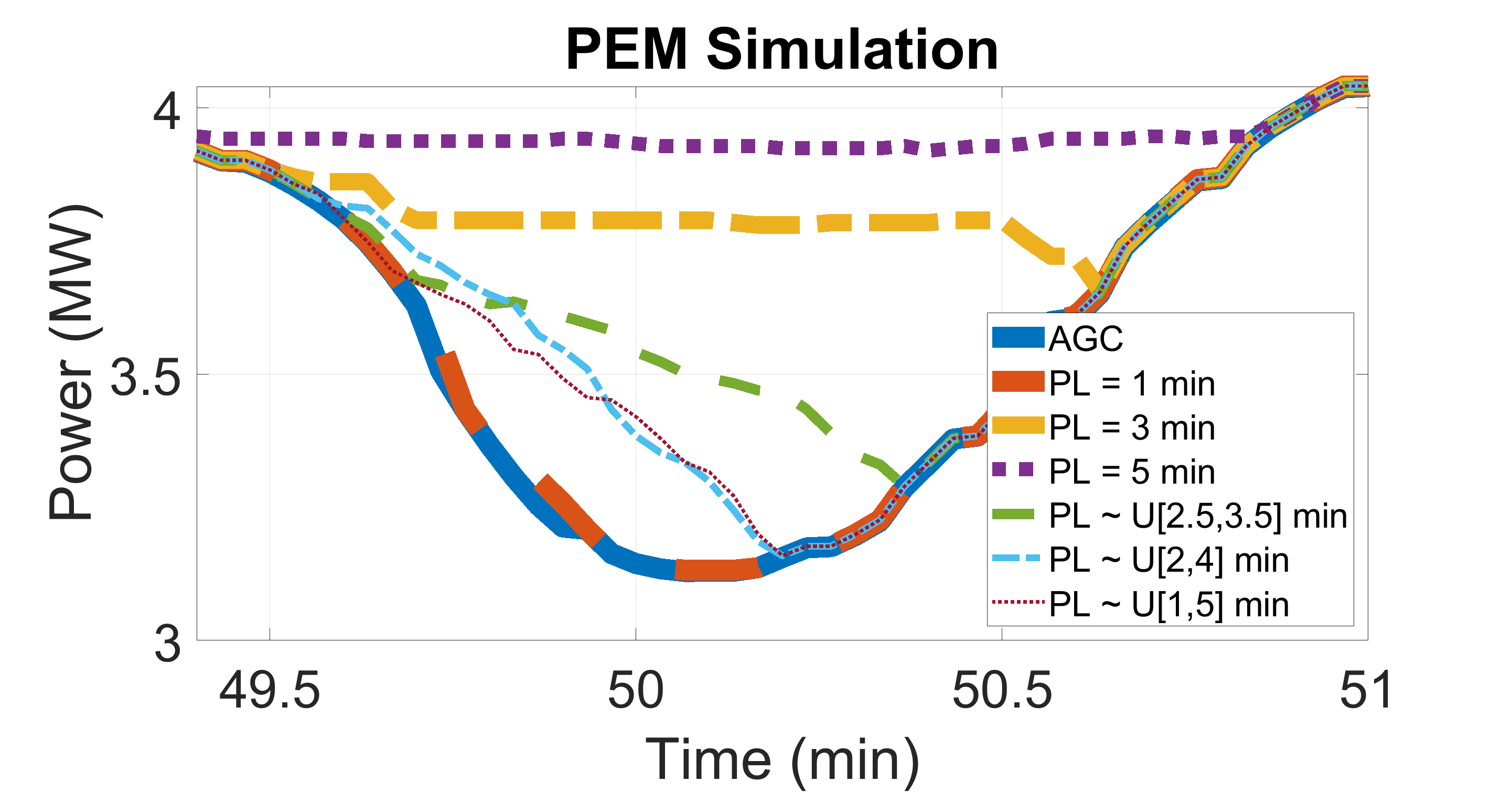}
    \caption{Effect of Packet Length on PEM Response}
    \label{fig:pkt_rnd}
\end{figure}

Consider the snippet of the AGC signal shown in Fig.~\ref{fig:pkt_rnd} and the response of PEM for different packet lengths. Recall that in PEM, the coordinator is not allowed to interrupt a packet once it has been accepted. Therefore, this effect is more prominent in the case of $5$ min packets (purple dotted line), where the coordinator is locked into requests accepted at and before minute $49.5$ in Fig.~\ref{fig:pkt_rnd} resulting in a larger tracking error as compared to the case of $3$ (yellow dash-dotted line) and $1$ min (thick red dashed line) packets.

\begin{figure}[t]
    \centering
    \includegraphics[width=1\columnwidth]{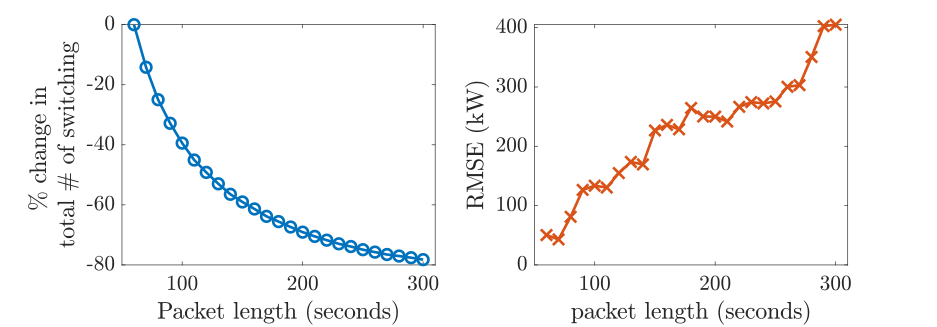}
    \caption{Effects of packet length on device cycling and tracking error. (Left) illustrates that increasing packet length reduces device cycling significantly. (Right) shows increasing tracking error with increasing packet length.}
    \label{fig:sw_no_rnd}
\end{figure}

Although tracking performance improves with shorter packets, however, it increases device cycling. A DER cycles between ON and OFF mode as its request is accepted and then the packet is subsequently consumed. DERs are usually limited by physical constraints that reduce their lifetime as cycling increases. Reducing packet length increases the number of times a DER transitions between ON and OFF modes, thereby increasing device cycling, as shown in Fig.~\ref{fig:sw_no_rnd} (left plot), where packet length is varied from $1$ min to $5$ mins. In Fig.~\ref{fig:sw_no_rnd} the y-axis is with respect to $1$ min packet length. Furthermore, the tracking RMSE increases as the packet length increases as expected since due to down-ramp limited response in case of longer packet lengths. Cycling can be included as a constraint in demand dispatch, which is especially important in the case of coordinating air-conditioners and HVAC thermostats~\cite{zhangTstatsACs2013,prabir:cycl}. The next subsection shows that introducing randomization on packet lengths surprisingly reduces cycling.

\subsection{Packet randomization}
Time-varying packet lengths are introduced in PEM logic by allowing each DER's requested packet length to vary according to a specific distribution. Particularly, this work focuses on uniform distribution $\mathcal{U}(\delta_\text{p} - \delta_\text{a},\delta_\text{p} + \delta_\text{a})$ where $\delta_\text{p}$ is the mean packet length and $\delta_\text{a}$ is the offset corresponding to the edges of the uniform distribution. 

From an implementation perspective, packet randomization can be realized either at the PEM coordinator or at the DER's local controller. In the first method, the coordinator assigns each DER whose request has been accepted a specific packet length by embedding this information in the response to DER's request. The coordinator assigns packet lengths to the requesting DER randomly, from a known distribution with bounded support.

For the second method, each DER's local PEM logic is modified so that the requested packet length is drawn from the uniform distribution and this information is embedded into the request message. Both of these methods improve PEM's tracking performance as discussed next.
\begin{figure}[t]
    \centering
    \includegraphics[width=\columnwidth]{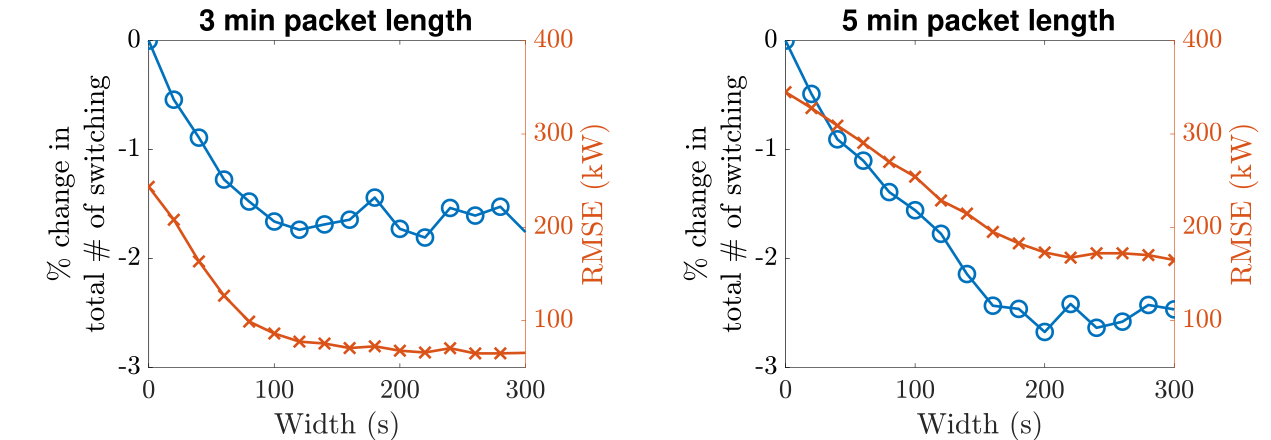}
    \caption{Effect of packet randomization on device cycling with mean $\delta_{\text{p}}=3$ min (left) and $\delta_{\text{p}}=5$ min (right). The packet length is drawn from a uniform distribution $\mathcal{U}[\delta_\text{p}-\delta_\text{a}, \delta_\text{p}+\delta_\text{a}]$ and the width is equal to to $2\delta_\text{a}$ which is the length of the interval corresponding to the uniform distribution. Also, zero width means no packet randomization.}
    \label{fig:pkt_rnd_3min_5min}
\end{figure} 

The response of the PEM system when packet length is drawn from a uniform distribution $\mathcal{U}[2.5, 3.5]$ min with a mean of $3$ min is shown in green in Fig.~\ref{fig:pkt_rnd}. Compared to the case when each packet is $3$ min long and without randomization, the tracking performance improves with randomization due to the presence of shorter packets that allows tighter tracking. Furthermore, it can be seen from Fig.~\ref{fig:pkt_rnd} that PEM's response in green dashed line, with packet length $~\mathcal{U}[2.5,3.5]$ min, is closer to the case when packets are $1$ min long. As the variance of the width of the uniform distribution increases to 2 min, with a mean of 3 min, the tracking performance improves (blue dash-dotted line), although after that the performance improvement saturates.
\begin{remark}
Although the packet length of requests made by DERs in Fig.~\ref{fig:pkt_rnd} is randomized according to a uniform distribution at each time step, this scheme can also be implemented by generating a string of requests drawn from the same uniform distribution for the entire duration at the start of the simulation and for all DERs. The DERs can then cycle deterministically through the string of packet lengths while requesting the PEM coordinator resulting in similar tracking performance.
\end{remark}
\subsection{Device cycling in time-varying packets}
Introducing randomization in packet lengths not only improves tracking error but also has the added benefit of reducing device cycling. Tracking error improves due to the availability of shorter packets. Consider the case of packets distributed according to $\mathcal{U}[2.5, 3.5]$. Compared to the case with only a single packet length of $3$ min, randomization allows packets less than $3$ min, resulting in better tracking of down-ramps. Surprisingly, cycling also decreases due to the availability of a range of packets in the interval $[2.5, 3.5]$ min which has the combined effect of reducing the overall number of accepted requests and hence the number of times a device transitions between ON and OFF states. Fig.~\ref{fig:pkt_rnd_3min_5min} shows that for two different mean packet lengths, the device cycling decreases with time-varying packets as the width of the uniform distribution increases compared to the case without time-varying packets. The plots in Fig.~\ref{fig:pkt_rnd_3min_5min} compare a single realization of the PEM system for each of the mean packet lengths. The next subsection quantifies the tracking improvements due to packet randomization by conducting sensitivity studies.

\subsection{Tracking Performance}
\begin{figure}[t]
    \centering
    \includegraphics[width=0.5\columnwidth]{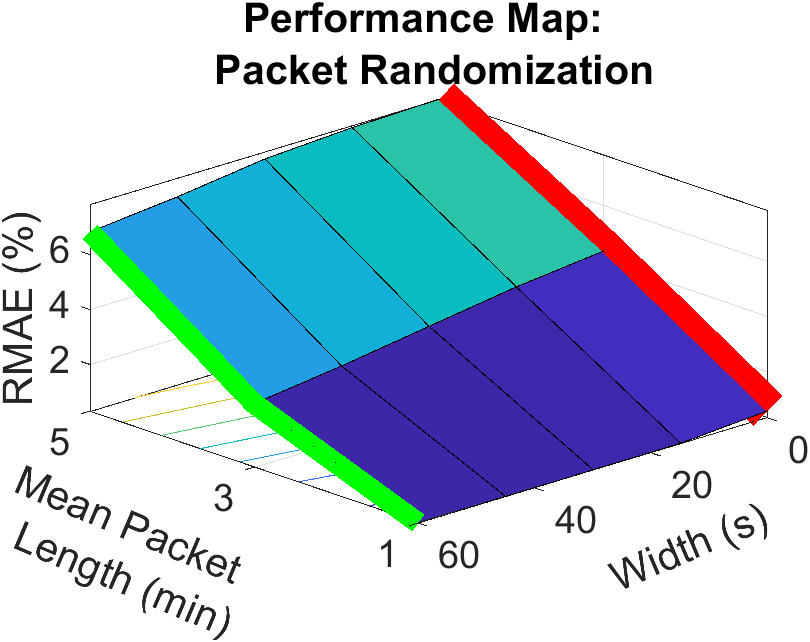}
    \caption{Performance Map with Packet Randomization. 
    Packet lengths are drawn from a uniform distribution $\mathcal{U}[\delta_\text{p}-\delta_\text{a}, \delta_\text{p}+\delta_\text{a}]$ where $\delta_\text{p}$ is the mean packet length and $2\delta_\text{a}$ is the width of the uniform distribution interval.
    }
    \label{fig:pkt_rnd_pmap}
\end{figure}

Simulations were performed with the same twelve 1-h AGC signals considered in Section \ref{mpcsim} but with packet randomization and no MPC. The packet lengths were drawn from a uniform distribution. The results are shown in Fig. \ref{fig:pkt_rnd_pmap}. It can be seen that there is a mean improvement in RMAE of around 1\% when the mean packet length is 5 min and the width of the uniform distribution is 1 min, i.e., the packet lengths $\sim \mathcal{U}[4.5, 5.5]$ min. The mean improvement in RMAE is more for 3 min for a width of 1 min, i.e., around 2\%. This is because in that case, smaller packets are available for tracking. Moreover, the performance improves uniformly with the width of the uniform distribution. Comparing Fig. \ref{fig:pkt_rnd_pmap} with Fig. \ref{fig:pmaplp} or \ref{fig:pmapqp}, we can see that the performance improvement is lesser with packet randomization with respect to MPC-based precompensator for the same packet length and 1 min-width.
Note that the red line of Fig. \ref{fig:pkt_rnd_pmap} (representing plain PEM) does not match exactly that of Fig. \ref{fig:pmaplp} (which corresponds to an MPC prediction horizon of 2 s) at 5 min packet length, due to a slight plant/model mismatch that increases with packet length.

\section{Conclusion}\label{sec:Con}
In this paper, packetized energy management has been augmented with a model predictive control framework for providing fast grid services. 
The MPC-based precompensator uses a linearized VB model for PEM and a prediction model for AGC and is shown to improve performance by about 4\% in RMAE, 10\% in RRMSE, and  15\% in PJM performance scores when the forecast is perfect. Essentially, the effect of MPC on PEM (with perfect forecast) is similar to reducing packet length by two minutes. This is important, as it reduces the switching and communication burden on the coordination scheme, which is beneficial from an implementation perspective. Furthermore, a passive time-varying packet-length mechanism is also developed for PEM. Simulation results show that time-varying packet length improves tracking performance by 1-2\%, due to the presence of shorter packets. Future work includes developing a VB model for PEM for randomized packet lengths, developing improved AGC forecasting tools, and combining the methods of MPC and packet randomization to achieve further improved tracking performance.

\vspace{-5pt}
\appendix[PJM Performance Scores]
We provide here the definitions of the PJM performance scores that are depicted in Fig. \ref{pjmscores}. They were obtained from the Excel sheet provided in \cite{pjmxls}. In the following, TREG refers to the total regulation capacity of the fleet, AREG the PJM-assigned regulation capacity, $r$ the reference AGC signal, \textcolor{black}{$r_{\textrm{max}}$ and $r_{\textrm{min}}$ the maximum and minimum values of $r$ respectively, }$r_{0}$ the economic basepoint, $\overline{r_{0}}$ the ramp-limited economic basepoint, $\textrm{RR}_{10}$ the ramp-limit, URES the unit resource response, UREG the unit-specific share of the regulation signal (i.e., the ideal response expected from the resource), and $y$ the output of the resource. The step size for computing PJM performance scores is considered 10 s, so that if a smaller sampling time is used, as done in the simulations in this paper, averaging may be used to combine the data points. 

\subsubsection{Precision Score}
The PJM precision score is given by:
\begin{equation*}
\textrm{Precision Score}=\frac{1}{360}\sum_{k=1}^{360}p_{\textrm{prec}}\left[k\right]
\end{equation*}
where:
\begin{align*}
    p_{\textrm{prec}}\left[k\right]&=\begin{cases}
0, & \textrm{TREG}\left[k\right]=0\\
\tilde{p}_{\textrm{prec}}\left[k\right], & \textrm{TREG}\left[k\right]\neq0
\end{cases}\\
\tilde{p}_{\textrm{prec}}\left[k\right]&=\begin{cases}
0, & \max\left\{ r\left[k\right],\ldots r\left[k+60\right]\right\} \\
 & =\textrm{min}\left\{ r\left[k\right],\ldots r\left[k+60\right]\right\}\\
 \widehat{p}_{\textrm{prec}}\left[k\right], & \textrm{otherwise}
\end{cases}\\
\widehat{p}_{\textrm{prec}}\left[k\right]&=\textrm{min}\left\{ \textrm{max}\left(0,1-\frac{\left|\textrm{URES}\left[k\right]-\textrm{UREG}\left[k\right]\right|}{\overline{\textrm{UREG}}\left[k\right]}\right),1\right\}\\
&\textrm{if }\textrm{min}\left\{ \textrm{AREG}\left[k\right],\ldots\textrm{AREG}\left[k+30\right]\right\} \neq0
\end{align*}
\begin{align*}
\textrm{URES}\left[k\right]&=y\left[k\right]-\overline{r_{0}}[k]\\
\textrm{UREG}\left[k\right]&=\begin{cases}
\frac{\textrm{AREG}\left[k\right]}{\textrm{TREG}\left[k\right]}\hat{r}[k], & \textrm{TREG}\left[k\right]\neq0\\
0, & \textrm{TREG}\left[k\right]=0
\end{cases}\\
\hat{r}[k] &= 2\frac{r[k]-r_{0}[k]}{r_{\textrm{max}}-r_{\textrm{min}}}\\
\overline{\textrm{UREG}}\left[k\right]&=\frac{1}{240}\sum_{i=0}^{239}\left|\textrm{UREG}\left[k+i\right]\right|\\
\overline{r_{0}}\left[k\right]	&=\begin{cases}
r_{0}\left[k\right], & \left|r_{0}\left[k\right]-\right.\\
 & \left.\overline{r_{0}}\left[k-1\right]\right|<\textrm{RR}_{10}\left[k\right]\\
\overline{r_{0}}\left[k-1\right]\\
+\textrm{sgn}\left(r_{0}\left[k\right]\right.\\
\left.-\overline{r_{0}}\left[k-1\right]\right)\textrm{RR}_{10}\left[k\right], & \textrm{otherwise}
\end{cases}
\end{align*}
and sgn is the signum function.

\subsubsection{Accuracy Score}
The accuracy score is given by:
\begin{gather*}
\textrm{Accuracy Score}=\frac{1}{360}\sum_{k=1}^{360}p_{\textrm{acc}}\left[k\right]
\end{gather*}where:
\textcolor{black}{
\begin{align*}p_{\textrm{acc}}\left[k\right] & =\begin{cases}
0, & \textrm{TREG}\left[k\right]\neq0\\
\widehat{p}_{\textrm{acc}}\left[k\right], & \textrm{otherwise}
\end{cases}\\
\widehat{p}_{\textrm{acc}}\left[k\right] & =\begin{cases}
0, & \max\left\{ r\left[k\right],\ldots r\left[k+60\right]\right\} \\
 & =\textrm{min}\left\{ r\left[k\right],\ldots r\left[k+60\right]\right\} \\
\widetilde{\rho}\left[n\left[k\right],k\right], & \textrm{otherwise}
\end{cases}\\
\widetilde{\rho}\left[m,k\right] & =\begin{cases}
\rho\left[m,k\right], & \hat{x}[k]\geq0.05\\
\mu\left[m,k\right], & \hat{x}[k]<0.05
\end{cases}\\
x[k] & =\begin{cases}
0, & \textrm{TREG}\left[k\right]\neq0\\
1, & \frac{\hat{r}[k]}{\textrm{TREG}\left[k\right]}>1\\
-1, & \frac{\hat{r}[k]}{\textrm{TREG}\left[k\right]}<-1\\
\frac{\hat{r}[k]}{\textrm{TREG}\left[k\right]}, & \textrm{otherwise}
\end{cases}\\
\overline{x}[k] & =\frac{1}{31}\sum_{j=0}^{30}x[k+j],\quad\overline{k}[k]=\frac{1}{31}\sum_{j=0}^{30}\left(k+j\right)\\
\hat{x}[k] & =\frac{1}{\sqrt{30}}\sqrt{\sum_{j=0}^{30}\left(x[k+j]-\overline{x}[k]\right)^{2}}\\
\rho\left[m,k\right] & =\frac{\sum_{j=0}^{30}v_{1}\left[k,j\right]v_{2}\left[m,k,j\right]}{\sqrt{\sum_{j=0}^{30}v_{1}\left[k,j\right]^{2}\sum_{j=0}^{30}v_{2}\left[m,k,j\right]^{2}}}\\
v_{1}\left[k,j\right] & =\textrm{UREG}\left[k+j\right]-\overline{v}_{1}[k]\\
v_{2}\left[m,k,j\right] & =\textrm{URES}\left[k+m+j\right]-\overline{v}_{2}[m,k]\\
\overline{v}_{1}[k] & =\frac{1}{31}\sum_{j=0}^{30}\textrm{UREG}\left[k+j\right]\\
\overline{v}_{2}[m,k] & =\frac{1}{31}\sum_{j=0}^{30}\textrm{URES}\left[k+m+j\right]\\
\mu\left[m,k\right] & =1-\left|\widetilde{\mu}\left[k\right]-\hat{\mu}\left[m,k\right]\right|\\
\widetilde{\mu}\left[k\right] & =\frac{\sum_{j=0}^{30}\left(k+j-\overline{k}[k]\right)\left(x[k+j]-\overline{x}[k]\right)}{\sum_{j=0}^{30}\left(k+j-\overline{k}[k]\right)^{2}}\\
\hat{\mu}\left[m,k\right] & =\frac{\sum_{j=0}^{30}\left(j-15\right)v_{3}[m,k,j]}{2480}\\
v_{3}[m,k,j] & =\textrm{URES}[k+m+j]-\overline{\textrm{URES}}[m,k]\\
\overline{\textrm{URES}}[m,k] & =\frac{1}{31}\sum_{j=0}^{30}\textrm{URES}[k+m+j]\\
n\left[k\right] & =\underset{m\in\left\{ 0,...,30\right\} }{\textrm{argmax}}\left\{ \frac{1}{3}\textrm{max}\left(\textrm{min}\left(1,\widetilde{\rho}\left[m,k\right]\right),0\right)\right.\\
 & \left.+\frac{1}{3}\widehat{p}_{\textrm{del}}\left[m\right]\right\} \\
\widehat{p}_{\textrm{del}}\left[m\right] & =\textrm{min}\left(1-\frac{m-1}{30},1\right)
\end{align*}}

\subsubsection{Delay Score}
The delay score is given by:
\begin{equation*}
\textrm{Delay Score}=\frac{1}{360}\sum_{k=1}^{360}p_{\textrm{del}}\left[k\right]
\end{equation*}
where:
\begin{align*}
    p_{\textrm{del}}\left[k\right]&=\begin{cases}
0, & \widehat{p}_{\textrm{acc}}\left[k\right]=0\\
\widehat{p}_{\textrm{del}}\left[n\left[k\right]\right], & \textrm{otherwise}
\end{cases}\\
&\textrm{if }\textrm{min}\left\{ \textrm{AREG}\left[k\right],\ldots\textrm{AREG}\left[k+30\right]\right\} \neq0
\end{align*}
and $\widehat{p}_{\textrm{del}}[m]$, $n[k]$, and $\widehat{p}_{\textrm{acc}}[k]$ are as used in the definition of the accuracy score.

\subsubsection{Composite Score}
Finally, the composite score is the average of the accuracy, delay and precision scores:
\begin{gather*}
    \textrm{Composite Score}=\frac{1}{3}\left(\textrm{Accuracy Score}+\textrm{Delay Score}\right.\\+\left.\textrm{Precision Score}\right)
\end{gather*}

\section*{Acknowledgment}
The authors would like to thank Dr. Mehdi Firouznia, formerly a postdoctoral scholar at The University of Vermont, for initial experiments on packet randomization. 

\ifCLASSOPTIONcaptionsoff
  \newpage
\fi

\bibliographystyle{IEEEtran}
\bibliography{ref}

\begin{thebibliography}{10}
\providecommand{\url}[1]{#1}
\csname url@samestyle\endcsname
\providecommand{\newblock}{\relax}
\providecommand{\bibinfo}[2]{#2}
\providecommand{\BIBentrySTDinterwordspacing}{\spaceskip=0pt\relax}
\providecommand{\BIBentryALTinterwordstretchfactor}{4}
\providecommand{\BIBentryALTinterwordspacing}{\spaceskip=\fontdimen2\font plus
\BIBentryALTinterwordstretchfactor\fontdimen3\font minus
  \fontdimen4\font\relax}
\providecommand{\BIBforeignlanguage}[2]{{%
\expandafter\ifx\csname l@#1\endcsname\relax
\typeout{** WARNING: IEEEtran.bst: No hyphenation pattern has been}%
\typeout{** loaded for the language `#1'. Using the pattern for}%
\typeout{** the default language instead.}%
\else
\language=\csname l@#1\endcsname
\fi
#2}}
\providecommand{\BIBdecl}{\relax}
\BIBdecl

\bibitem{Morgan:1979}
M.~Morgan and S.~Talukdar, ``Electric power load management: Some technical,
  economic, regulatory and social issues,'' \emph{Proceedings of the IEEE},
  vol.~67, no.~2, pp. 241 -- 312, 1979.

\bibitem{schweppe1980homeostatic}
F.~C. Schweppe, R.~D. Tabors, J.~L. Kirtley, H.~R. Outhred, F.~H. Pickel, and
  A.~J. Cox, ``Homeostatic utility control,'' \emph{IEEE Trans.~Power Apparatus
  and Systems}, no.~3, pp. 1151--1163, 1980.

\bibitem{Brooks:DemandDispatch}
A.~{Brooks}, E.~{Lu}, D.~{Reicher}, C.~{Spirakis}, and B.~{Weihl}, ``Demand
  dispatch,'' \emph{IEEE Power and Energy Magazine}, vol.~8, no.~3, pp. 20--29,
  May 2010.

\bibitem{Callaway:2011wq}
D.~S. Callaway and I.~A. Hiskens, ``{Achieving Controllability of Electric
  Loads},'' \emph{Proceedings of the IEEE}, vol.~99, no.~1, pp. 184--199, Jan.
  2011.

\bibitem{Kolter:PESGM2016}
X.~Zhang, G.~Hug, J.~Z. Kolter, and I.~Harjunkoski, ``Model predictive control
  of industrial loads and energy storage for demand response,'' in \emph{2016
  IEEE Power and Energy Society General Meeting (PESGM)}, 2016, pp. 1--5.

\bibitem{PJM_AGC_url}
{PJM}, ``{PJM Ancillary Services},''
  \url{https://www.pjm.com/markets-and-operations/ancillary-services.aspx},
  {Last Updated}: {2020-07-06}.

\bibitem{pjmmanual}
Dispatch, \emph{PJM Manual 12: Balancing Operations, Revision 42}.\hskip 1em
  plus 0.5em minus 0.4em\relax PJM, 2021.

\bibitem{PJM_pay4perf}
{Andy Ott, PJM}, ``{ Frequency Regulation Market Pay for Performance},''
  \url{https://www.caiso.com/Documents/FrequencyRegulationMarketPay\_Per
  formance-PJMPresentationbyAndyOtt.pdf}, {Last Updated}: 2011-04.

\bibitem{Callaway2009ECM}
D.~S. Callaway, ``Tapping the energy storage potential in electric loads to
  deliver load following and regulation, with application to wind energy,''
  \emph{Energy Conversion and Management}, vol.~50, no.~5, pp. 1389 -- 1400,
  2009.

\bibitem{hao:tpwrst}
H.~Hao, B.~M. Sanandaji, K.~Poolla, and T.~L. Vincent, ``Aggregate flexibility
  of thermostatically controlled loads,'' \emph{IEEE Transactions on Power
  Systems}, vol.~30, no.~1, pp. 189--198, 2015.

\bibitem{Mathieu:2013TPWRS}
J.~L. Mathieu, S.~Koch, and D.~S. Callaway, ``{State Estimation and Control of
  Electric Loads to Manage Real-Time Energy Imbalance},'' \emph{IEEE
  Transactions on Power Systems}, vol.~28, no.~1, pp. 430--440, 2013.

\bibitem{Luminita2014IFAC}
L.~C. Totu and R.~Wisniewski, ``Demand response of thermostatic loads by
  optimized switching-fraction broadcast,'' \emph{IFAC Proceedings Volumes},
  vol.~47, no.~3, pp. 9956 -- 9961, 2014, 19th IFAC World Congress.

\bibitem{abate:tcst}
S.~Esmaeil Zadeh~Soudjani and A.~Abate, ``Aggregation and control of
  populations of thermostatically controlled loads by formal abstractions,''
  \emph{IEEE Transactions on Control Systems Technology}, vol.~23, no.~3, pp.
  975--990, 2015.

\bibitem{Meyn2015TAC}
S.~P. {Meyn}, P.~{Barooah}, A.~{Bušić}, Y.~{Chen}, and J.~{Ehren},
  ``Ancillary service to the grid using intelligent deferrable loads,''
  \emph{IEEE Transactions on Automatic Control}, vol.~60, no.~11, pp.
  2847--2862, Nov 2015.

\bibitem{kundu:fitness}
S.~P. Nandanoori, S.~Kundu, D.~Vrabie, K.~Kalsi, and J.~Lian, ``Prioritized
  threshold allocation for distributed frequency response,'' in \emph{2018 IEEE
  Conference on Control Technology and Applications (CCTA)}, 2018, pp.
  237--244.

\bibitem{kundu:ai}
I.~Chakraborty, S.~P. Nandanoori, S.~Kundu, and K.~Kalsi, ``Stochastic virtual
  battery modeling of uncertain electrical loads using variational
  autoencoder*,'' in \emph{2020 American Control Conference (ACC)}, 2020, pp.
  1305--1310.

\bibitem{hassan:privacyaware}
A.~Hassan, D.~Deka, and Y.~Dvorkin, ``Privacy-aware load ensemble control: A
  linearly-solvable mdp approach,'' 2021.

\bibitem{emiliano:multi_period_OPF}
E.~Benenati, M.~Colombino, and E.~Dall’Anese, ``A tractable formulation for
  multi-period linearized optimal power flow in presence of thermostatically
  controlled loads,'' in \emph{2019 IEEE 58th Conference on Decision and
  Control (CDC)}, 2019, pp. 4189--4194.

\bibitem{Garcia:privacy_DER}
M.~Zholbaryssov, C.~N. Hadjicostis, and A.~D. Dominguez-Garcia,
  ``Privacy-preserving distributed coordination of distributed energy
  resources,'' in \emph{2020 59th IEEE Conference on Decision and Control
  (CDC)}, 2020, pp. 4689--4696.

\bibitem{kleissl:tsg}
T.~Anderson, M.~Muralidharan, P.~Srivastava, H.~V. Haghi, J.~Cortés,
  J.~Kleissl, S.~Martínez, and B.~Washom, ``Frequency regulation with
  heterogeneous energy resources: A realization using distributed control,''
  \emph{IEEE Transactions on Smart Grid}, pp. 1--1, 2021.

\bibitem{Lian:transactive}
Y.~Lu, J.~Lian, and M.~Zhu, ``Privacy-preserving transactive energy system,''
  in \emph{2020 American Control Conference (ACC)}, 2020, pp. 3005--3010.

\bibitem{DuffautEspinosa:2020tpwrs}
L.~A. Duffaut~Espinosa and M.~Almassalkhi, ``A packetized energy management
  macromodel with quality of service guarantees for demand-side resources.''
  \emph{IEEE Transactions on Power Systems}, vol.~35, no.~5, pp. 3660--3670,
  2020.

\bibitem{DuffautEspinosa:2020tcst}
L.~A. Duffaut~Espinosa, A.~Khurram, and M.~Almassalkhi, ``A packetized energy
  management macromodel with quality of service guarantees for demand-side
  resources.'' \emph{IEEE Transactions on Power Systems}, to appear.

\bibitem{fathy:tcst}
S.~Bashash and H.~K. Fathy, ``Modeling and control of aggregate air
  conditioning loads for robust renewable power management,'' \emph{IEEE
  Transactions on Control Systems Technology}, vol.~21, no.~4, pp. 1318--1327,
  2013.

\bibitem{angstoch}
D.~Angeli and P.-A. Kountouriotis, ``A stochastic approach to
  “dynamic-demand” refrigerator control,'' \emph{IEEE Transactions on
  Control Systems Technology}, vol.~20, no.~3, pp. 581--592, 2012.

\bibitem{tindeControl}
S.~H. Tindemans, V.~Trovato, and G.~Strbac, ``Decentralized control of
  thermostatic loads for flexible demand response,'' \emph{IEEE Transactions on
  Control Systems Technology}, vol.~23, no.~5, pp. 1685--1700, 2015.

\bibitem{Almassalkhi:2018IMA}
M.~Almassalkhi, L.~A. Duffaut~Espinosa, P.~D. Hines, J.~Frolik, S.~Paudyal, and
  M.~Amini, \emph{Asynchronous Coordination of Distributed Energy Resources
  with Packetized Energy Management}.\hskip 1em plus 0.5em minus 0.4em\relax
  New York, NY: Springer New York, 2018, pp. 333--361.

\bibitem{DuffautEspinosa:2018PSCC}
L.~A. {Duffaut Espinosa}, M.~Almassalkhi, P.~Hines, and J.~Frolik, ``System
  properties of packetized energy management for aggregated diverse
  resources,'' \emph{Power Systems Computation Conference}, June 2018.

\bibitem{DuffautEspinosa:2020cdc}
L.~A. Duffaut~Espinosa, A.~Khurram, and M.~Almassalkhi, ``A virtual battery
  model for packetized energy management,'' in \emph{59th IEEE Conference on
  Decision and Control (CDC)}, December 2020, pp. 42--48.

\bibitem{prabir:cycl}
A.~R. Coffman, N.~Cammardella, P.~Barooah, and S.~Meyn, ``Flexibility capacity
  of thermostatically controlled loads with cycling/lock-out constraints,'' in
  \emph{2020 American Control Conference (ACC)}, 2020, pp. 527--532.

\bibitem{koch2011modeling}
S.~Koch, J.~L. Mathieu, D.~S. Callaway \emph{et~al.}, ``Modeling and control of
  aggregated heterogeneous thermostatically controlled loads for ancillary
  services,'' in \emph{Proc. PSCC}.\hskip 1em plus 0.5em minus 0.4em\relax
  Citeseer, 2011, pp. 1--7.

\bibitem{khurram:phd}
\BIBentryALTinterwordspacing
A.~Khurram, ``\BIBforeignlanguage{English}{Modeling and control for packetized
  energy management},'' Ph.D. dissertation, 2021, copyright - Database
  copyright ProQuest LLC; ProQuest does not claim copyright in the individual
  underlying works; Last updated - 2021-11-27. [Online]. Available:
  \url{https://www.proquest.com/dissertations-theses/modeling-control-packetized-energy-management/docview/2594546885/se-2?accountid=14524}
\BIBentrySTDinterwordspacing

\bibitem{espinosa2020virtual}
L.~A.~D. Espinosa, A.~Khurram, and M.~R. Almassalkhi, ``A virtual battery model
  for packetized energy management,'' in \emph{2020 59th IEEE Conference on
  Decision and Control (CDC)}.\hskip 1em plus 0.5em minus 0.4em\relax IEEE,
  2020, pp. 42--48.

\bibitem{box2015time}
G.~E. Box, G.~M. Jenkins, G.~C. Reinsel, and G.~M. Ljung, \emph{Time series
  analysis: forecasting and control}.\hskip 1em plus 0.5em minus 0.4em\relax
  John Wiley \& Sons, 2015.

\bibitem{pfeifer1980identification}
P.~E. Pfeifer and S.~J. Deutsch, ``Identification and interpretation of first
  order space-time arma models,'' \emph{Technometrics}, vol.~22, no.~3, pp.
  397--408, 1980.

\bibitem{akaike1979bayesian}
H.~Akaike, ``A bayesian extension of the minimum aic procedure of
  autoregressive model fitting,'' \emph{Biometrika}, vol.~66, no.~2, pp.
  237--242, 1979.

\bibitem{gurobi}
\BIBentryALTinterwordspacing
{Gurobi Optimization, LLC}, ``{Gurobi Optimizer Reference Manual},'' 2021.
  [Online]. Available: \url{https://www.gurobi.com}
\BIBentrySTDinterwordspacing

\bibitem{pjmxls}
\BIBentryALTinterwordspacing
PJM, ``40-minute performance score template - updated to reflect august mrc
  changes xls 10.9.2,'' workbook in Microsoft Excel. [Online]. Available:
  \url{\url{https://www.pjm.com/-/media/markets-ops/ancillary/mkt-based-regulation/40-minute-performance-score-template-updated-to-reflect-august-mrc-changes.ashx}}
\BIBentrySTDinterwordspacing

\bibitem{zhangTstatsACs2013}
W.~Zhang, J.~Lian, C.-Y. Chang, and K.~Kalsi, ``Aggregated modeling and control
  of air conditioning loads for demand response,'' \emph{IEEE Transactions on
  Power Systems}, vol.~28, no.~4, pp. 4655--4664, 2013.

\end{thebibliography}

\end{document}